\documentclass[aps,prd,twocolumn,superscriptaddress]{revtex4-1}

\usepackage[pdftex]{graphicx}
\usepackage{amsmath}
\usepackage{amsfonts}
\usepackage{amssymb}

\begin{document}


\title{A method for high precision reconstruction of air shower Xmax using two-dimensional radio intensity profiles}




\author{S.~Buitink}
\affiliation{Department of Astrophysics/IMAPP, Radboud University Nijmegen, 6500 GL Nijmegen, The Netherlands}
\author{A.~Corstanje}
\affiliation{Department of Astrophysics/IMAPP, Radboud University Nijmegen, 6500 GL Nijmegen, The Netherlands}
\author{J.~E.~Enriquez}
\affiliation{Department of Astrophysics/IMAPP, Radboud University Nijmegen, 6500 GL Nijmegen, The Netherlands}
\author{H.~Falcke}
\affiliation{Department of Astrophysics/IMAPP, Radboud University Nijmegen, 6500 GL Nijmegen, The Netherlands}
\affiliation{Netherlands Institute for Radio Astronomy (ASTRON), Postbus 2, 7990 AA Dwingeloo, The Netherlands}
\affiliation{Nikhef, Science Park Amsterdam, 1098 XG Amsterdam, The Netherlands}
\affiliation{Max-Planck-Institut f\"{u}r Radioastronomie, Auf dem H\"ugel 69, 53121 Bonn, Germany}
\author{J.~R.~H\"orandel}
\affiliation{Department of Astrophysics/IMAPP, Radboud University Nijmegen, 6500 GL Nijmegen, The Netherlands}
\affiliation{Nikhef, Science Park Amsterdam, 1098 XG Amsterdam, The Netherlands}
\author{T.~Huege}
\affiliation{IKP, Karlsruhe Institute of Technology (KIT), Postfach 3640, 76021 Karlsruhe, Germany}
\author{A.~Nelles}
\affiliation{Department of Astrophysics/IMAPP, Radboud University Nijmegen, 6500 GL Nijmegen, The Netherlands}
\author{J.~P.~Rachen}
\affiliation{Department of Astrophysics/IMAPP, Radboud University Nijmegen, 6500 GL Nijmegen, The Netherlands}
\author{P.~Schellart}
\affiliation{Department of Astrophysics/IMAPP, Radboud University Nijmegen, 6500 GL Nijmegen, The Netherlands}
\author{O.~Scholten}
\affiliation{KVI CART, University of Groningen, 9747 AA Groningen, The Netherlands}
\author{S.~ter Veen}
\affiliation{Department of Astrophysics/IMAPP, Radboud University Nijmegen, 6500 GL Nijmegen, The Netherlands}
\author{S.~Thoudam}
\affiliation{Department of Astrophysics/IMAPP, Radboud University Nijmegen, 6500 GL Nijmegen, The Netherlands}
\author{T.~N.~G.~Trinh}
\affiliation{KVI CART, University of Groningen, 9747 AA Groningen, The Netherlands}


\date{\today}

\begin{abstract}
The mass composition of cosmic rays contains important clues about their origin. Accurate measurements are needed to resolve long-standing issues such as the transition from Galactic to extragalactic origin, and the nature of the cutoff observed at the highest energies. 
Composition can be studied by measuring the atmospheric depth of the shower maximum $X_\mathrm{max}$ of air showers generated by high-energy cosmic rays hitting the Earth's atmosphere.
We present a new method to reconstruct $X_\mathrm{max}$ based on radio measurements. The radio emission mechanism of air showers is a complex process that creates an asymmetric intensity pattern on the ground. The shape of this pattern strongly depends on the longitudinal development of the shower.
We reconstruct $X_\mathrm{max}$ by fitting two-dimensional intensity profiles, simulated with CoREAS, to data from the LOFAR radio telescope. In the dense LOFAR core, air showers are detected by hundreds of antennas simultaneously. The simulations fit the data very well, indicating that the radiation mechanism is now well-understood.
The typical uncertainty on the reconstruction of $X_\mathrm{max}$ for LOFAR showers is 17 g/cm$^2$.   
  
\end{abstract}


\maketitle

\section{Introduction}
High-energy cosmic rays are routinely measured by experiments around the world, yet there are still many urgent questions about their origin. Cosmic rays below 10$^{16}$ eV are expected to be of Galactic origin while the highest energy cosmic rays are likely to come from extragalactic sources. However, it is not known how and at what energy this transition takes place. Another important question is whether the cut-off at the end of the spectrum \cite{AugerCutoff, HiResCutoff} is due to interactions of cosmic ray protons with the cosmic microwave background \cite{Greisen, Zatsepin}, or marks the highest energy that cosmic accelerators can reach. Accurate measurements of the cosmic ray mass composition are needed to resolve these questions \cite{KU12}. In addition, a clean separation between protons and heavy nuclei above $6\times 10^{19}$~eV will greatly improve the search for a correlation between cosmic ray arrival directions and their sources \cite{AugerCorrelation}.

Several techniques exist to measure mass-sensitive shower parameters. Particle detector arrays like KASCADE-Grande \cite{KGrande} and IceTop \cite{IceTop} measure the electron-to-muon ratio of the secondary shower particles at ground level. This ratio depends on the mass of the primary particle, but also on the age of the shower, which makes the technique susceptible to shower-to-shower fluctuations. 

The mass composition can also be inferred from the distribution of $X_\mathrm{max}$, the atmospheric depth of the shower maximum. At the Pierre Auger observatory, the longitudinal shower profile is measured by observing the fluorescence light that is emitted by air molecules that were excited by shower particles \cite{AugerFD}. Alternatively, $X_\mathrm{max}$ can be inferred by measuring the optical Cherenkov light emitted by the shower particles with arrays like Tunka \cite{Tunka}. Both techniques require dark nights and are therefore severely limited in duty cycle (below 15\%). The fluorescence technique yields the best precision on $X_\mathrm{max}$ of $\sim 20$~g/cm$^2$.

Here we propose a new technique that has the same precision but has a duty cycle of almost 100\%, based on the radio emission produced by air showers. Like fluorescence light, the radio signal carries information of the complete longitudinal development of the shower \cite{lopes_muon}. It is therefore possible to reconstruct $X_\mathrm{max}$ from the radio signal \cite{lopes_ldf}. 
The method presented here requires measurements with many radio antennas simultaneously, in order to adequately sample the radiation profile.
We apply this technique to data from the LOFAR radio telescope \cite{vanHaarlem2013, Schellart2013, FG03}.

Our reconstruction technique requires precise simulations of the radio emission. The theory of air shower radio emission has developed rapidly in the last decade, and simulation codes based on different approaches are now converging towards similar results \cite{ModelConvergence}. In this work, we use the CoREAS simulation package \cite{CoREAS}, which is a plug-in for CORSIKA \cite{CORSIKA} that computes the radio pulse by adding the contribution of all individual electrons and positrons in the shower. We demonstrate that simulations are in excellent agreement with the LOFAR measurements, which sample the radio emission with hundreds of antennas per individual shower.

Section \ref{sec:lofar} gives a brief introduction to the LOFAR radio telescope array. In Section \ref{sec:simulations} we describe the simulation tools and settings. The new reconstruction technique is explained in Section \ref{sec:technique} and Section \ref{sec:systematics} contains a review of systematic effects in the reconstruction. Conclusions are given in the final Section.

\section{LOFAR}
\label{sec:lofar}
The Low Frequency Array (LOFAR) is a new-generation radio telescope constructed in the North of the Netherlands with satellite stations across Europe. It consists of thousands of dipole antennas, sensitive in the frequency range of $10-240$~MHz. LOFAR has a flexible design that allows many different observation modes, some of which can run simultaneously. The raw electromagnetic waveform as it is measured by the dipoles, is stored for five seconds on ring buffers for each active antenna.

\begin{figure}
\includegraphics[width=\linewidth]{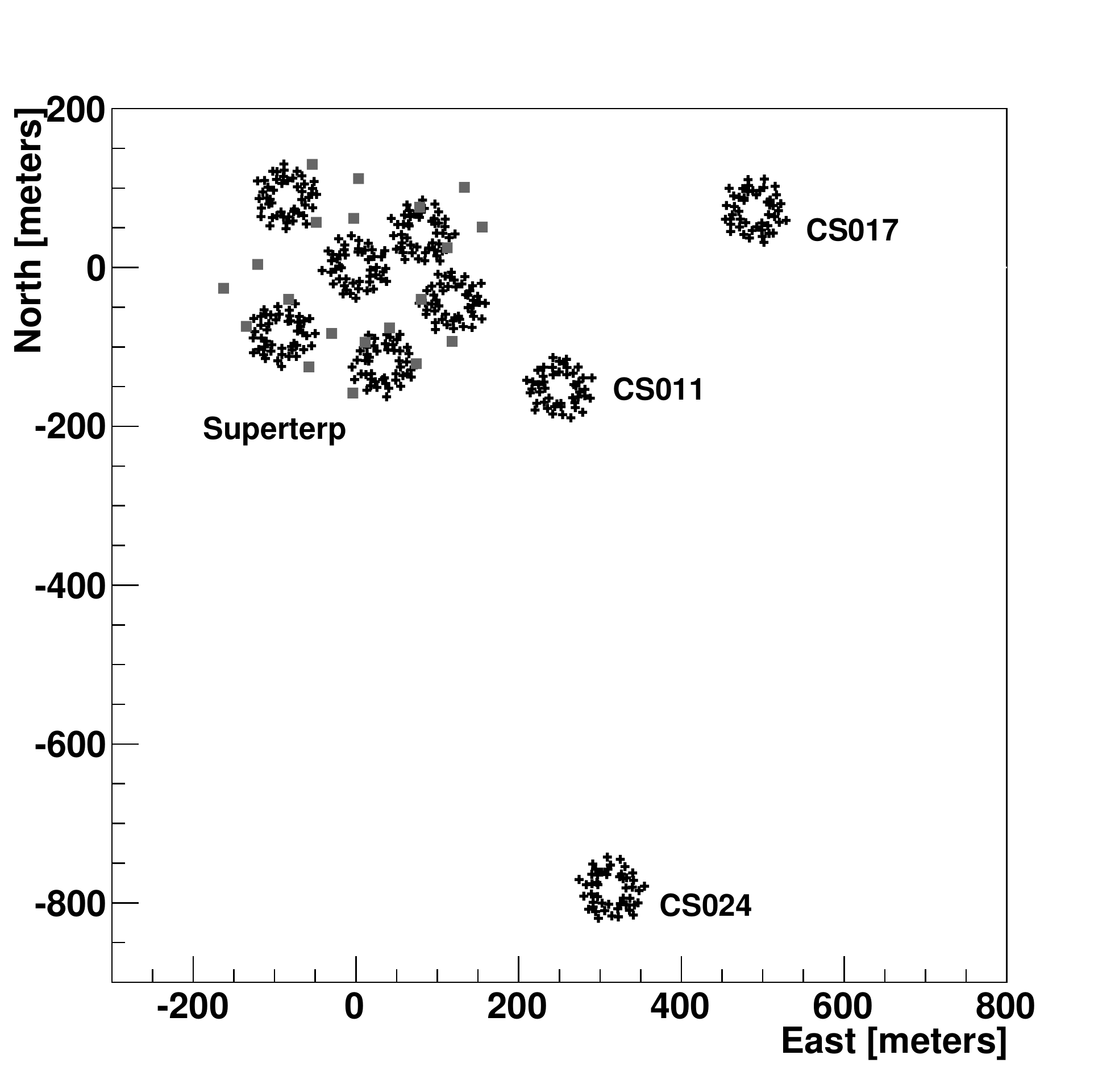}
\caption{\label{fig:layout} Part of the LOFAR core. The ``+'' signs indicate the position of antennas which are arranged in stations. Only the LBA antennas in the outer rings of each station are plotted. The inner part of the LBA stations and the HBA stations, which are not used in this study, are not included in the plot. The central six stations form the \emph{superterp}. Other core stations lie at increasingly larger distances from this cluster, three of which are visible in the map. The grey squares indicate the positions of LORA detectors.}
 \end{figure}
  
LOFAR is organized in stations, each containing 96 low band antennas (LBA; 10-90 MHz) and 48 high band antennas (HBA; 110-240 MHz). In the center of LOFAR, six of these stations are placed close together on a small artificial island, called the \emph{superterp} (see Fig.~\ref{fig:layout}). Other stations are placed around this island at increasing distances. A total of 24 stations form the LOFAR core which has a diameter of $\sim 2$~km. For cosmic-ray detection we focus on this dense core region.

The core is augmented with the LOFAR Radboud Air Shower Array (LORA) \cite{LORA} which comprises twenty particle detectors. They cover the superterp area and are an essential part of the cosmic-ray measurement capability. LORA detects air showers above $10^{16}$~eV and provides a trigger for the radio antennas. It reconstructs the arrival direction, core position and gives a first energy estimate of the shower. Radio pulses can unambiguously be associated to air showers when their arrival time, and direction coincides with the LORA reconstruction.

When a LORA trigger is received, a 2 ms trace is read out from the ring buffers of all active antennas and stored for offline processing \cite{Schellart2013}. A radio signal is typically found in hundreds of antennas. Depending on the mode of the current astronomical observations, either LBA or HBA data is available. Since HBA data is more challenging to analyse \cite{LOFAR_HBA}, we only consider LBA data in this study. 

\section{Radio emission simulations}
\label{sec:simulations}

\subsection{Emission mechanism}
\label{sec:mechanism}
The dominant component of the radio emission of air showers is driven by the geomagnetic field \cite{KL66, FG03}. The electrons and positrons are deflected in opposite direction by the Lorenz force and their drift creates a current perpendicular to the shower axis. As the shower develops, this current first grows and then decays, producing radio emission. The radiation is linearly polarized in the ${\bf v} \times {\bf B}$ direction, where ${\bf v}$ is the velocity of the shower front, and ${\bf B}$ is the Earth's magnetic field. 

As the shower propagates through the atmosphere, it also develops an excess of negative charge, due to
knock-out electrons from atmospheric molecules joining the shower, and the annihilation of positrons.
The growth and subsequent decay of the charge excess gives rise to a secondary emission component \cite{Askaryan}. Charge excess radiation is also linearly polarized, but in a different direction, pointing radially outwards from the shower axis. The relative contribution of charge excess to the total emission depends on the geometry (angle to the geomagnetic field, zenith angle, observer position, etc.) and has typical values of 5-20\% of the total pulse amplitude \cite{VriesScholten, AERA, Schellart2014}.

Because geomagnetic and charge excess radiation are polarized in different directions, the total emission is found by adding these contribution vectorially. The radiation profile can be easiest understood when it is plotted in the shower plane, with axes in the direction of ${\bf v} \times {\bf B}$ and ${\bf v} \times ({\bf v} \times {\bf B})$ (see Fig.~\ref{fig:twodepths_view}). In this frame, there is total constructive interference between the two components along the ${\bf v} \times {\bf B}$ axis in the positive direction, while the interference is most destructive in the negative direction along the same axis. Along the ${\bf v} \times ({\bf v} \times {\bf B})$ axis the two components are polarized orthogonally and add in quadrature. 

The resulting interference pattern is not rotationally symmetric and is typically bean-shaped. Evidently, the lateral distribution of the radio pulse power is not a one-dimensional function of distance to shower axis. The radio profile can only be accurately described in two dimensions.   

The bulk of the shower particles is confined to the shower front, a thin disc that travels towards the Earth at relativistic speed. For wavelengths exceeding the disc thickness (several meters) the radiation will be coherent (up to $\sim 100$~MHz). However, the propagation of the radio emission in a dielectric medium (air has an index of refraction of $n \approx 1.0003$ at sea level) produces Cherenkov-like effects which must be included to properly describe the radiation \cite{devries}. At the Cherenkov angle, the radio pulse is compressed and the emission is coherent up to GHz frequencies. 

\subsection{Experimental status}
Radio pulses from air showers were already detected in the 1960s \cite{Jelley}, but progress halted due to hardware limitations and a lack of proper understanding of the radiation mechanism. It was also feared that atmospheric electric fields would have a significant and unpredictable effect on the pulse strength. 

In the last decade, the interest in radio detection was revived and pursued with modern electronics \cite{Renaissance}. LOPES demonstrated that the emission mechanism is coherent and dominantly geomagnetic in nature \cite{LOPES05}, and that only the strong electric fields inside thunderstorms can significantly affect the radiation \cite{Buitink07}. 

Proof for a contribution of charge excess was found by CODALEMA \cite{codalema} and later by polarization studies of AERA \cite{AERA} and LOFAR \cite{Schellart2014}. CROME \cite{CROME} and ANITA \cite{ANITA} detected GHz emission at the Cherenkov angle, which can be interpreted as relativistic compression of the air shower radio emission. Full Cherenkov rings were found by LOFAR in the 110-190~MHz range \cite{LOFAR_HBA}. 

It has been shown by LOPES that the lateral distribution of the radio signal can be used to determine $X_\textrm{max}$ \cite{lopes_ldf}. Their approach was based on a one-dimensional approximation of the radio profile,
which yields a reconstruction resolution of 50 g/cm$^2$ for simulations, and 95 g/cm$^2$ for LOPES data.
Here we show that a much better resolution can be achieved by using a two-dimensional profile, and that
modern radio simulation codes like CoREAS can accurately predict the complete radiation profile as measured by LOFAR with hundreds of antennas per individual event, providing further proof that the emission mechanism is now understood to very high detail.

\subsection{CoREAS and CORSIKA}
In this work we use the radio simulation code CoREAS \cite{CoREAS} which is a plugin for the particle simulation code CORSIKA \cite{CORSIKA}. CoREAS is based on a microscopic description of the radiation mechanism, i.e.\ it computes the contributions of each electron and positron in the shower based on the `end-point' formalism \cite{endpoint}. In this formalism, the radiation produced by the acceleration of a particle at the start and end point of a particle track is calculated from first principles. By summing the contributions of all particle tracks the total emission can be calculated without making any assumptions on the type of radiation. In other words, while the radiation is best understood when explained in terms of geomagnetic and charge excess contributions, CoREAS does not simulate these components separately. Instead, it produces the complete radiation field that is generated by the distribution of charged particles simulated by CORSIKA. 

We use CORSIKA 7.400, with hadronic interaction models FLUKA 2011.2b \cite{fluka} and QGSJETII.04 \cite{qgsjet}. A comparison to other interaction models is made in Sec.~\ref{hadronic}. Thinning is applied at a level of $10^{-6}$ with optimized weight limitation \cite{kobal}.

A GEANT4 \cite{GEANT4} simulation of the LORA detectors \cite{LORA} is used to convert the CORSIKA particle output into the deposited energy as a function of distance to the shower axis. The particle lateral distribution function and radio profile will be fitted to the data simultaneously.   

\subsection{Two-dimensional profiles}
As explained above, the radiation profile is not rotationally symmetric and can only be accurately described by a two-dimensional intensity map. The CoREAS code computes the radio pulse for specific observer positions on the ground. A LOFAR antenna model is applied to these pulses to simulate  measured waveforms. This includes applying a frequency filter (10-90 MHz) and downsampling of the signal to 200 Msamples/s. In this analysis we use the pulse power integrated over a 55 ns time window centered at the pulse peak.

In order to derive a two-dimensional map, we run simulations for 160 ground positions and reconstruct the full profile by interpolation. For robust interpolation it is necessary to choose these positions such that they cover the locations where the interference between the two radiation component reaches its minimum and maximum. This is achieved by defining a star-shaped pattern in the shower plane with eight arms, two of which are aligned with the ${\bf v} \times {\bf B}$ axis, and projecting it onto the ground. Each arm contains 20 antennas, with a spacing of 25~m in the shower plane.

\begin{figure}
\includegraphics[width=\linewidth, trim= 0 30 70 30]{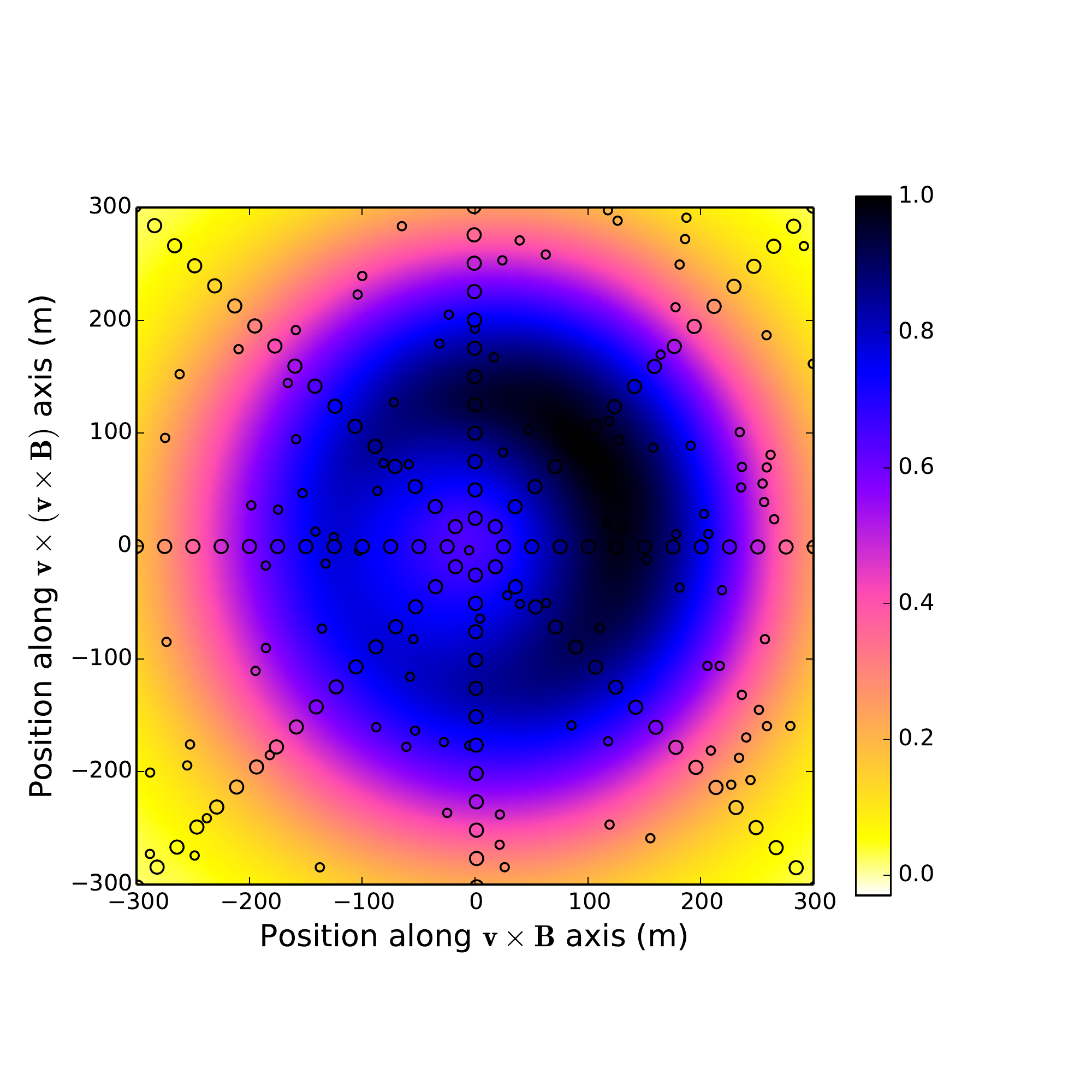}
\caption{\label{fig:interpolation} Two-dimensional profile of the received power integrated over a 55 ns window. The large circles indicate the positions for which CoREAS simulations have been generated, while the full map in the background is created by interpolation. The smaller circles indicate test positions for which simulated values are compared to interpolated values.}
 \end{figure}

In Fig.~\ref{fig:interpolation} the interpolation technique is demonstrated. The simulated antenna positions are marked by large circles, arranged in a regular pattern. Their color reflects the total power of the pulse as received by the antenna. The interpolated radio map is indicated by the background colors. By design, the interpolated map exactly matches the values at the simulated positions. To probe the error introduced by the interpolation, we run the same simulation for two hundred additional random positions, indicated by the smaller circles. The difference between the simulated and interpolated total power is always smaller than 2.5\% of the maximum power. This error is smaller than typical uncertainties in measured power for observations with LOFAR.
 
\begin{figure*}
\includegraphics[width=0.45\linewidth]{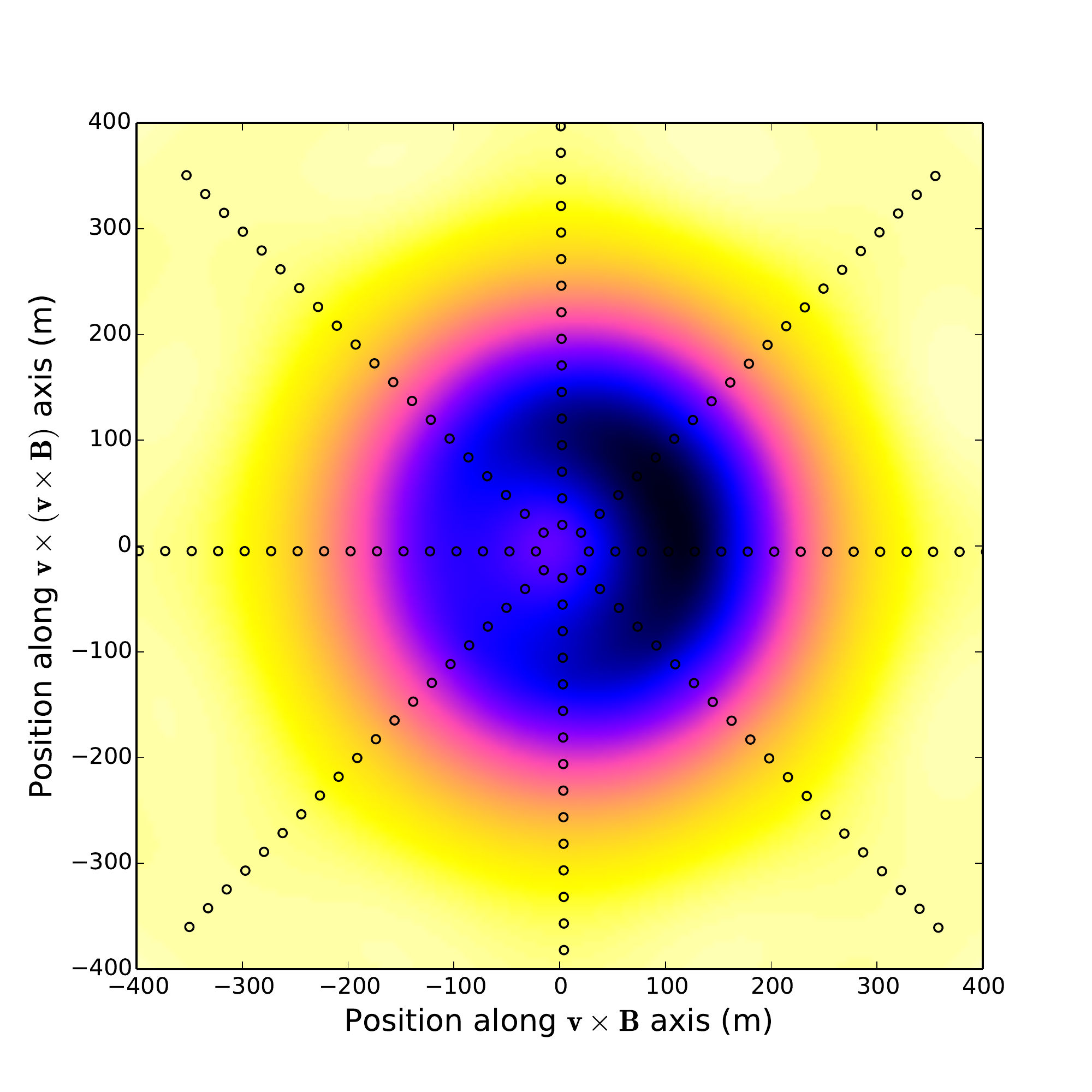}
\includegraphics[width=0.45\linewidth]{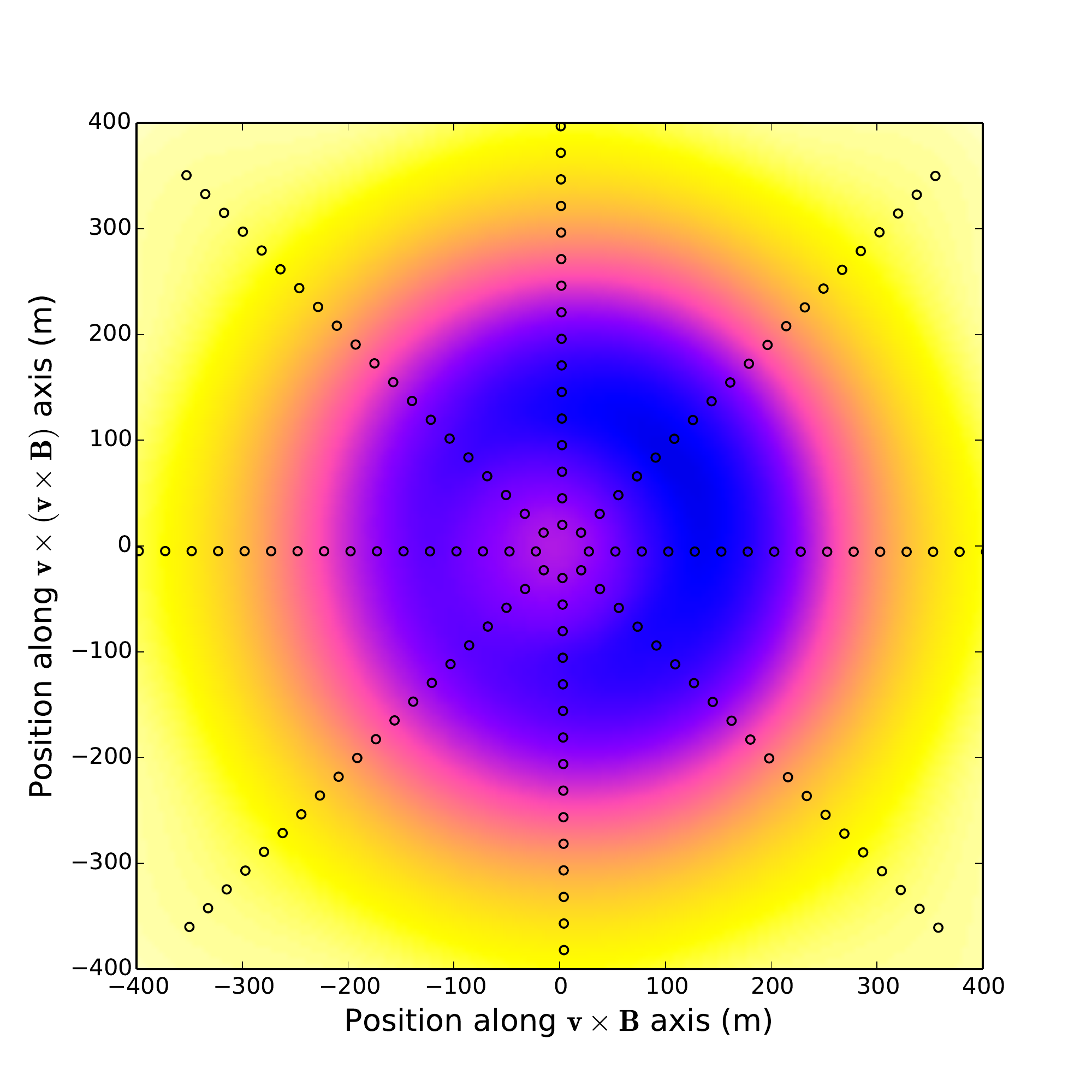}
\includegraphics[width=0.035\linewidth, trim= 10 -33 10 0]{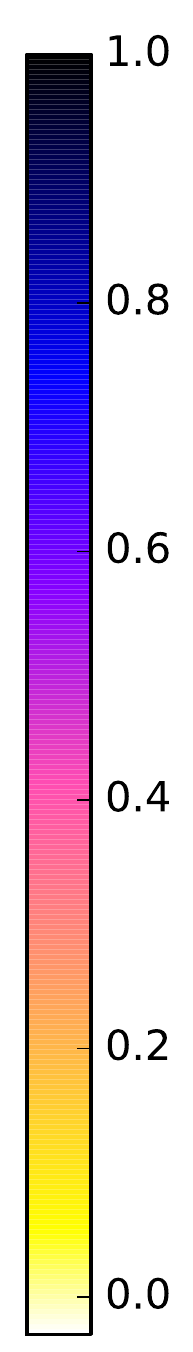}
\caption{\label{fig:twodepths_view} Radiation profiles of a proton shower with $X_\textrm{max}=794$ g/cm$^2$ (left panel) and an iron shower with $X_\textrm{max}=573$ g/cm$^2$ (right panel). Both showers have an energy of $2.3 \times 10^8$ GeV and a zenith angle of 49 degrees. The circles indicate the positions that have been simulated. The full background map is created by interpolation.}
\end{figure*}

\begin{figure}
\includegraphics[width=\linewidth]{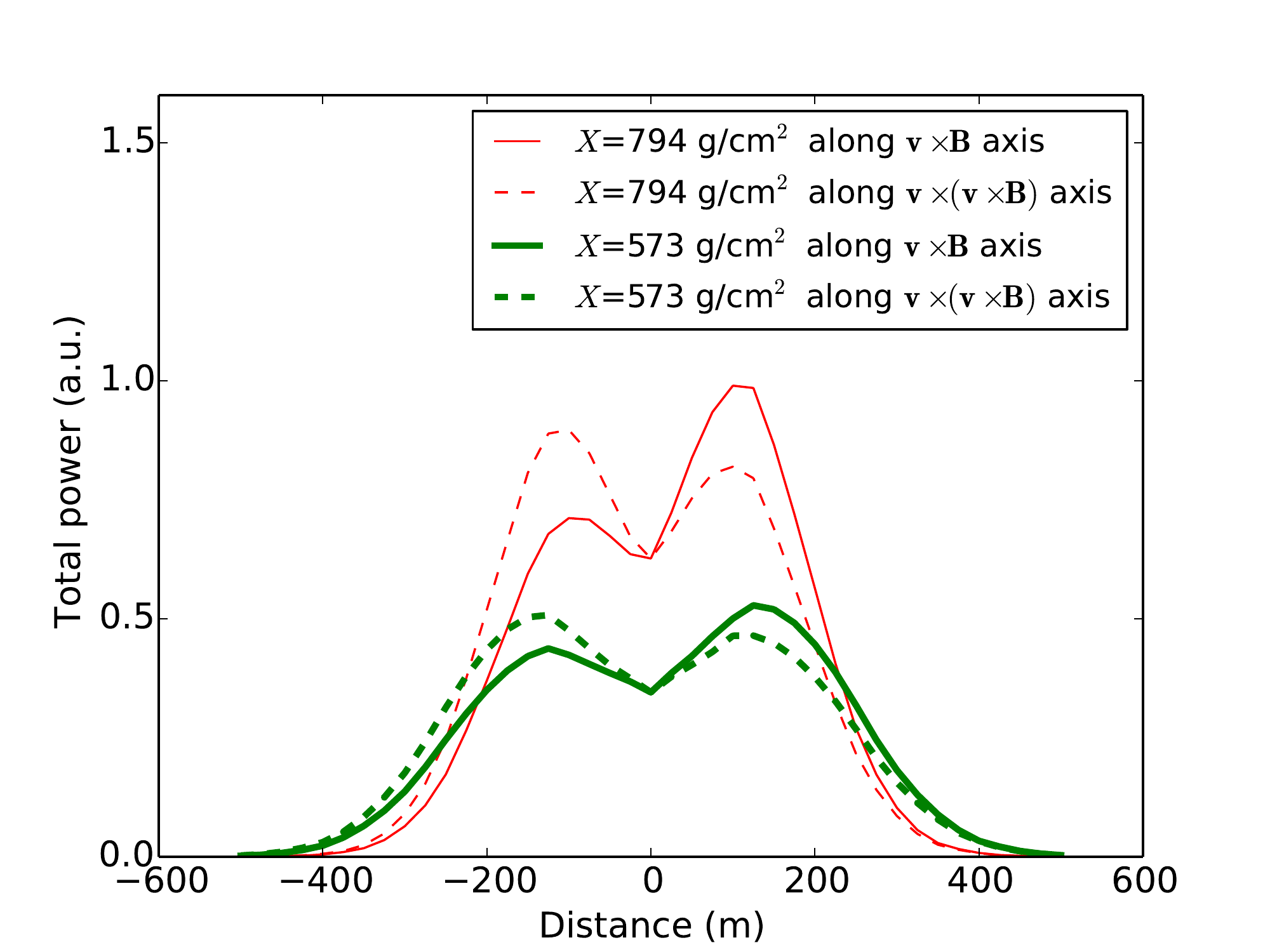}
\caption{\label{fig:twodepths} The pulse power as a function of position along the axes for the proton (red,thin) and iron (green, thick) showers shown in Fig.~\ref{fig:twodepths_view}.}
 \end{figure}

Fig.~\ref{fig:twodepths_view} shows the radio profiles for two showers that have been simulated with CoREAS. Both showers have a zenith angle of 49 degrees, an azimuth angle of 171 degrees (i.e.\ coming from the North-West), and an energy of $2.3 
\times 10^8$ GeV. The left panel displays the radiation profile of a proton shower that has penetrated deeply into the atmosphere, while the right panel shows the profile of a much shallower iron shower. In Fig.~\ref{fig:twodepths} the one-dimensional lateral distribution of the power is plotted for observers along the ${\bf v} \times {\bf B}$ and ${\bf v} \times ({\bf v} \times {\bf B})$ axes.  
 
Obviously, the profiles are very different. A general feature is that the power falls of more rapidly with distance for deeper showers, which is clearly seen in Fig.~\ref{fig:twodepths}. The $X_\textrm{max}$ reconstruction method used for LOPES is based on this feature \cite{lopes_ldf, HUE08}. 

However, there are also more subtle differences, like the amount of asymmetry and the position where the radiation reaches its maximum value. The method described in this paper makes use of all features of the radiation pattern by fitting the complete two-dimensional profile instead of a one-dimensional approximation.

The asymmetry along the ${\bf v} \times {\bf B}$ axis can be understood as the effect of the charge excess component.
It is interesting to note that there also exists some asymmetry along the ${\bf v} \times ({\bf v} \times {\bf B})$ axis. This is not expected from radiation physics reviewed in Section~\ref{sec:mechanism}. Indeed, when we plot the total `physical' pulse power as predicted by CoREAS there is no asymmetry along the ${\bf v} \times ({\bf v} \times {\bf B})$ axis. It only appears once the antenna response model is applied to the simulated pulses to calculate the total received power. The reason for this is that for observers at different locations the radiation has a different polarization. Since the antenna gain depends on the polarization, the power received by the antenna can be different even when the original pulse power is the same. In other words: different antennas pick up a different fraction of the total pulse power.

From Figs.~\ref{fig:twodepths_view} and \ref{fig:twodepths} it is clear that the shape of the radiation profile strongly depends on the atmospheric depth of the shower maximum $X_\textrm{max}$.
However, it is reasonable to assume that other variations in the longitudinal and lateral distribution of the shower also have an influence on the radiation pattern. Below we will demonstrate that the patterns are much more sensitive to $X_\textrm{max}$ than any other features of the shower development that vary from shower to shower. They do, however, limit the accuracy of the determination of $X_\textrm{max}$.

\subsection{Simulation set for LOFAR}
We have developed a reconstruction technique in which simulated two-dimensional radio profiles are fitted to data. We run dedicated simulations for each shower detected by LOFAR. 
The shower arrival direction is reconstructed based on the arrival time of the radio pulses at all antennas \cite{Schellart2013}. An energy estimate is provided by a LORA shower reconstruction \cite{LORA}. Since the shower core is often located outside the LORA array, this estimate is not accurate, and a better energy reconstruction is done at a later stage in the analysis. The core position itself is not needed as input for the simulation, since we use the star-shaped pattern of observer positions described above, instead of actual antenna positions.

For each shower in the set we run 25 proton showers and 15 iron showers. With this amount of showers we obtain a large range of $X_\textrm{max}$-values that reflects the natural spread.

\section{Hybrid reconstruction technique}
\label{sec:technique}
\subsection{Fit Procedure}
For each simulation we fit the two-dimensional radiation map and the one-dimensional particle lateral distribution function to the data simultaneously, by minimizing:
\begin{eqnarray}
\label{eq:fit}
\chi^2 = \sum_{\mathrm{antennas}} \left( \frac{P_{\mathrm{ant}} - f_r^2 P_{\mathrm{sim}}(x_{\mathrm{ant}}-x_0,y_{\mathrm{ant}}-y_0)}{\sigma_{\mathrm{ant}} }\right)^{2}  \nonumber \\
 +\sum_{\substack{\textrm{particle} \\ \textrm{detectors}}} \left( \frac{d_{\mathrm{det}} - f_p d_{\mathrm{\mathrm{sim}}}(x_{\mathrm{det}}-x_0,y_{\mathrm{det}}-y_0)}{\sigma_{\mathrm{det}} }\right)^{2},
\end{eqnarray}
where $P_{\mathrm{ant}}$ is the measured power integrated over a 55 ns window at an antenna at location $(x_{\mathrm{ant}},y_{\mathrm{ant}})$ with noise level $\sigma_{\mathrm{ant}}$,  $P_{\mathrm{sim}}$ is the simulated power, $d_{\mathrm{det}}$ is the deposited energy as measured by a LORA detector at location $(x_{\mathrm{det}},y_{\mathrm{det}})$ with noise $\sigma_{\mathrm{det}}$, and $d_{\mathrm{sim}}$ is the simulated deposited energy.
The fit contains four free parameters, two of which describe the location of the shower axis $(x_0,y_0)$. A scaling parameter $f_p$ for the particle lateral distribution function is needed to correct the energy scale, while a scaling parameter for the radio power $f_r^2$ is needed because the antennas do not yet have an absolute calibration. 
The radio power is approximately proportional to the square of the cosmic ray primary energy \cite{Huege05}, so both $f_r$ and $f_p$ scale linearly with energy. 

\subsection{Application to LOFAR data}
\begin{figure*}
\includegraphics[width=0.4\linewidth]{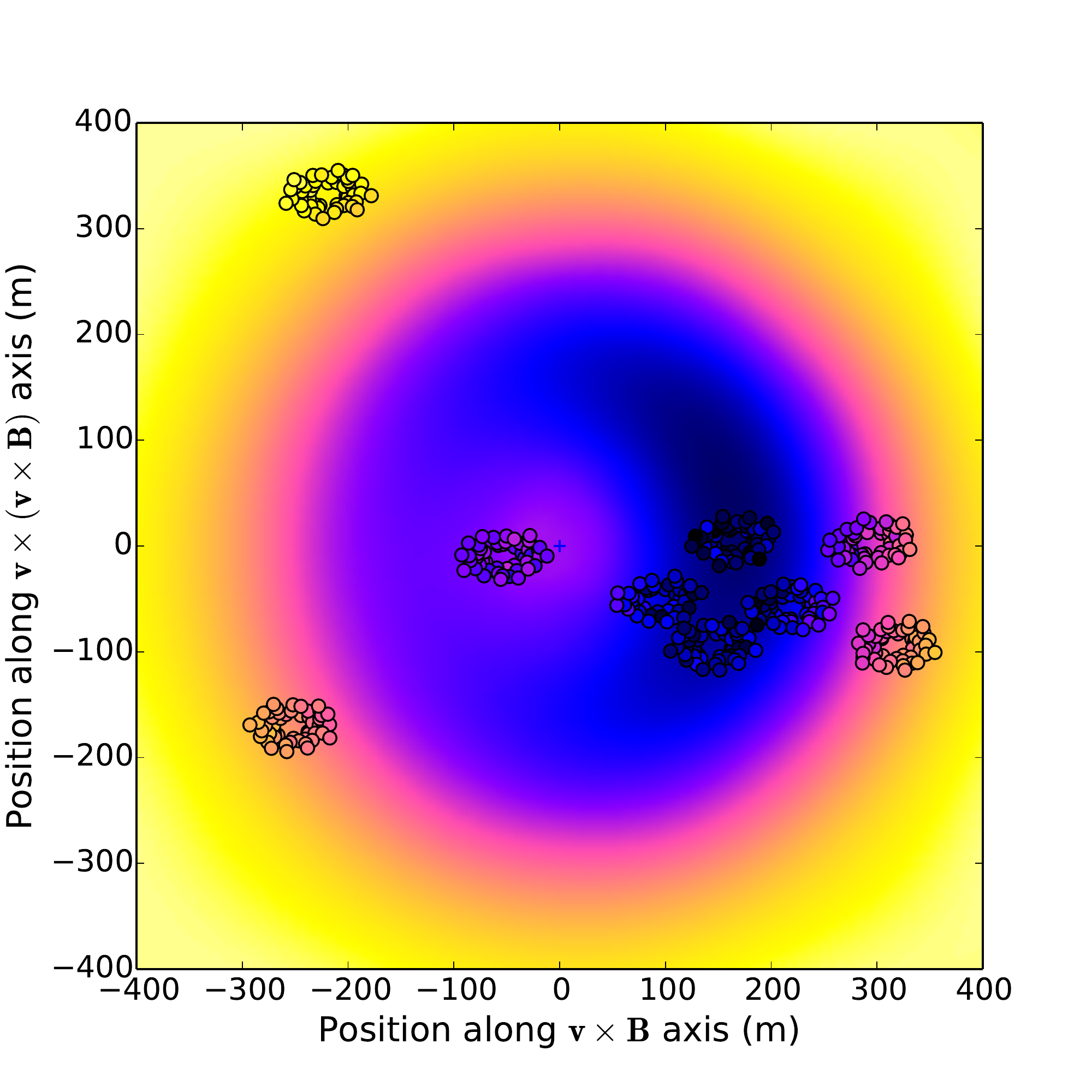}
\includegraphics[width=0.52\linewidth, trim= 0 -0.2cm 0 0]{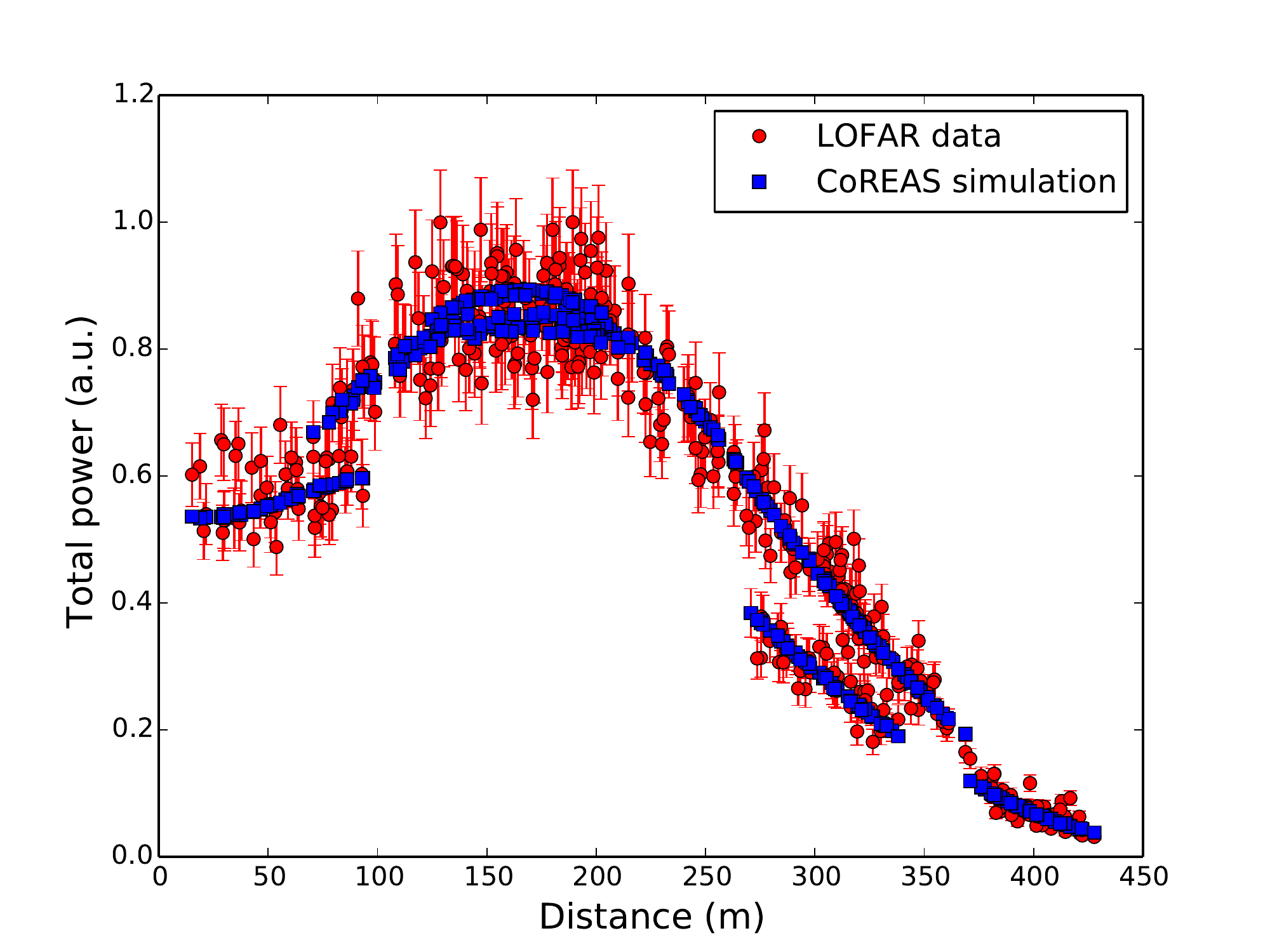}
\includegraphics[width=0.4\linewidth]{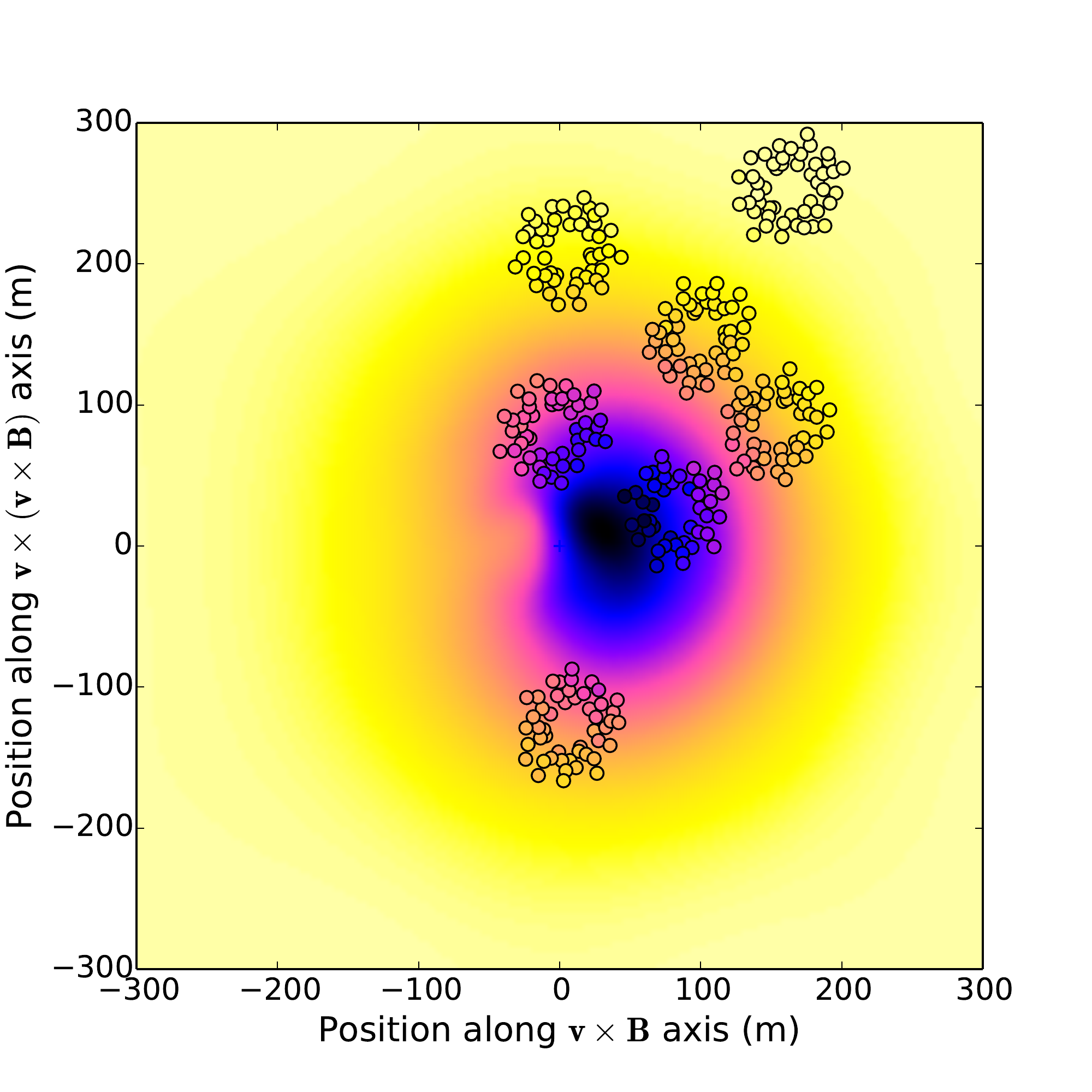}
\includegraphics[width=0.52\linewidth, trim= 0 -0.2cm 0 0]{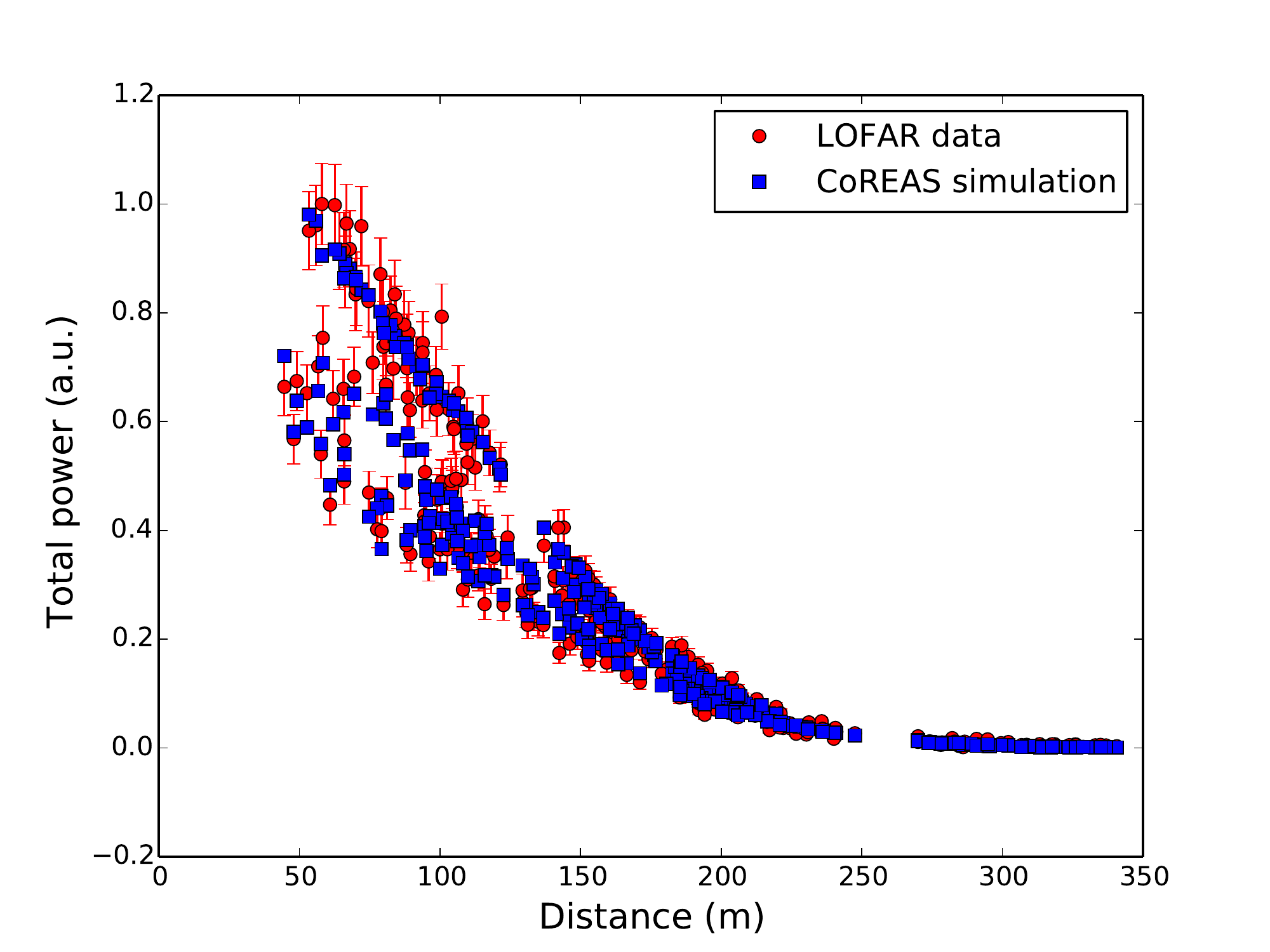}
\includegraphics[width=0.4\linewidth]{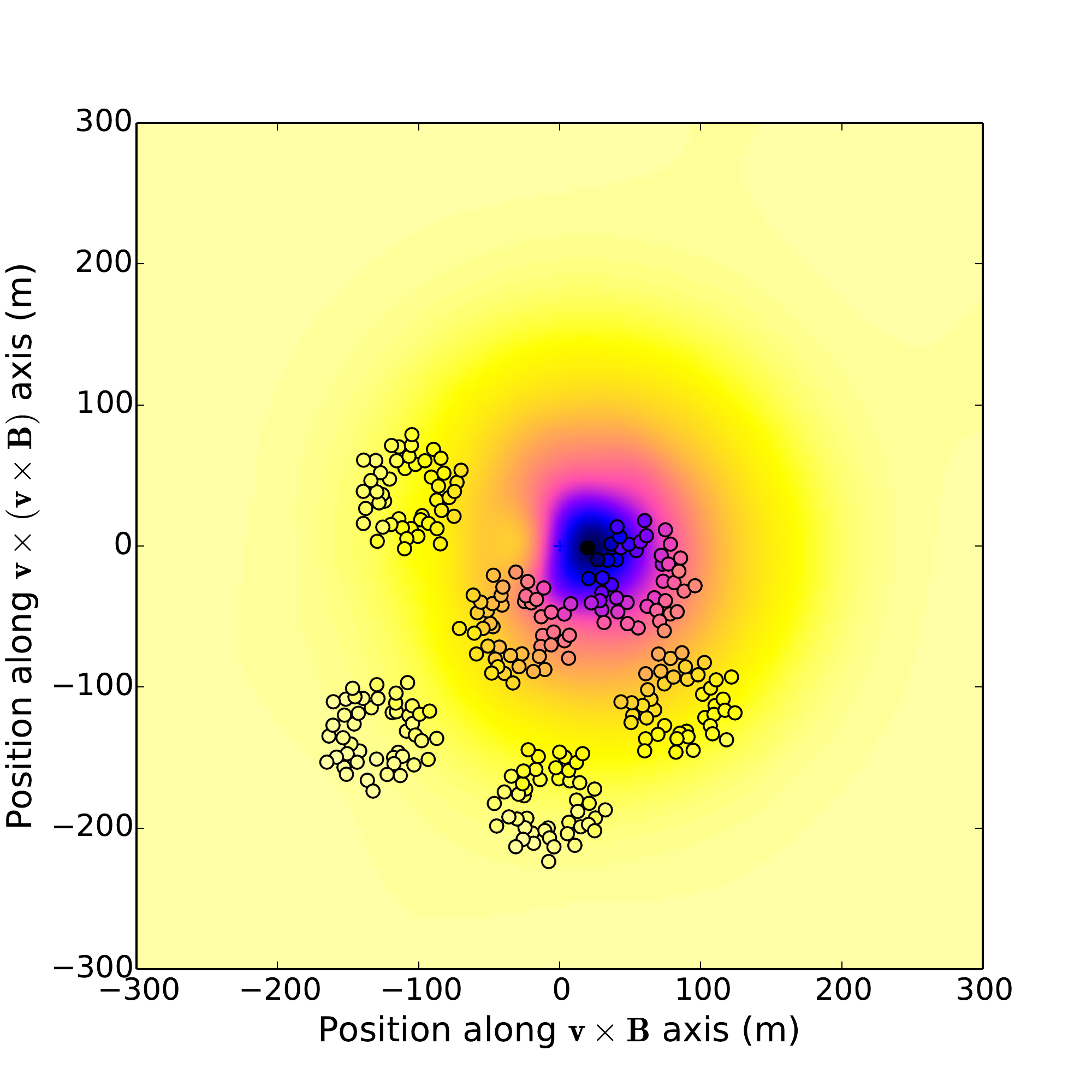}
\includegraphics[width=0.52\linewidth, trim= 0 -0.2cm 0 0]{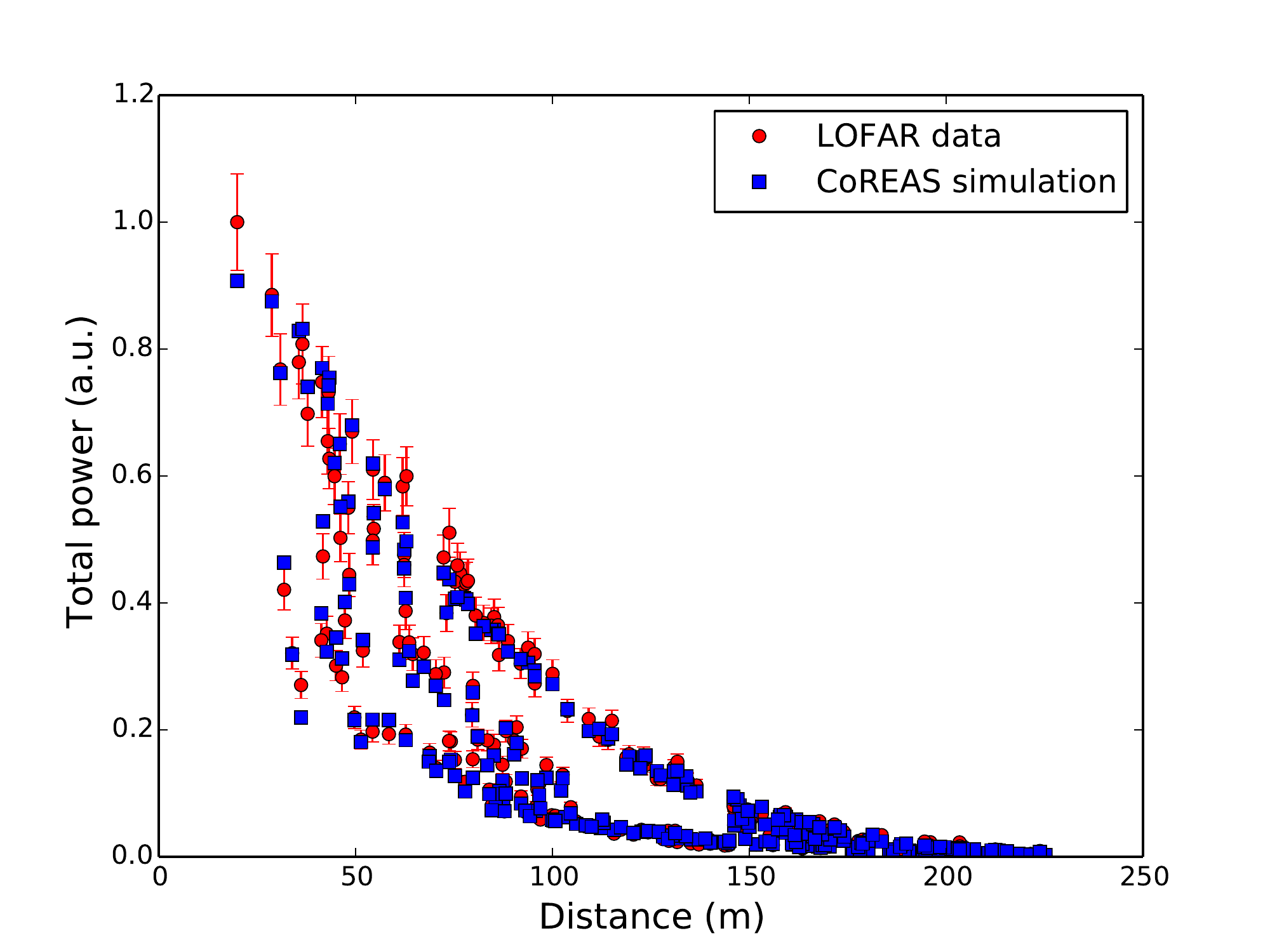}
\caption{\label{fig:events}Two-dimensional radio air shower reconstructions. The measured power for three different showers (top, middle, bottom) is fitted to a simulated radio map (left panels). The one-dimensional lateral distribution functions (right panels) are not single-valued functions of distance to the shower axis.}
\end{figure*}

Figure~\ref{fig:events} contains the fit results for three different showers. The left panels display the radiation pattern in the shower plane. The background colors indicate the simulated radio map, while the circles indicate the measurements of the LOFAR antennas. The color of the circle represents the received power at the antenna, so the fit is of high quality when the colors of the circles blend into the background. Note that the antennas are grouped in rings: the LOFAR stations. All these showers have been detected by all six superterp stations. In addition, emission from the shower in the top row was also found in three other core stations (CS024, CS011, CS017 in Fig.~\ref{fig:layout}), and the the shower in the middle row was detected by one additional station outside the superterp (CS011). For very inclined events the antenna rings appear flattened in the event display because of the projection onto the shower plane. As explained above, for each detected shower a total of forty simulations is produced. The results shown here are for the simulation that yielded the fit with the lowest $\chi^2$.

The right hand panels of Fig.~\ref{fig:events} show the same result in one dimension. For each antenna, the measured and simulated power is plotted as a function of distance to the shower axis. From all three examples it is clear that the radio power is not a single-valued function of distance. Some distribution functions contain oval structures (middle and bottom row) that are reminiscent of the ring structure of the LOFAR stations. From the shower in the top row, it is clear that the position where the radio power reaches its maximum can be far from the shower axis (in this case $\sim$150\ m). As explained in Sec.~\ref{sec:mechanism} this is due to the interference between geomagnetic and charge excess radiation, and Cherenkov-like propagation effects due to the non-unity index of refraction of air.   

\begin{figure}
\includegraphics[width=0.8\linewidth]{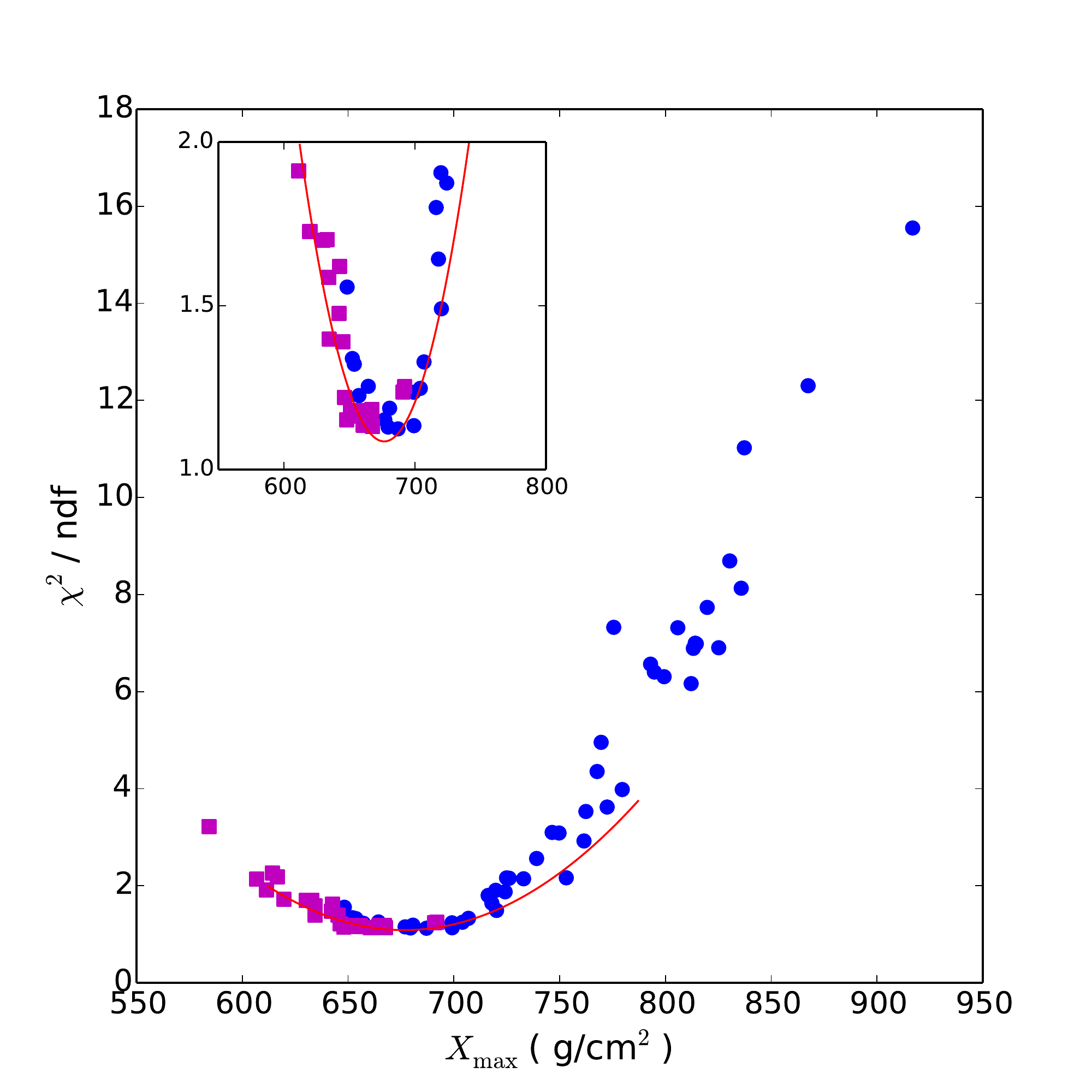}
\includegraphics[width=0.8\linewidth]{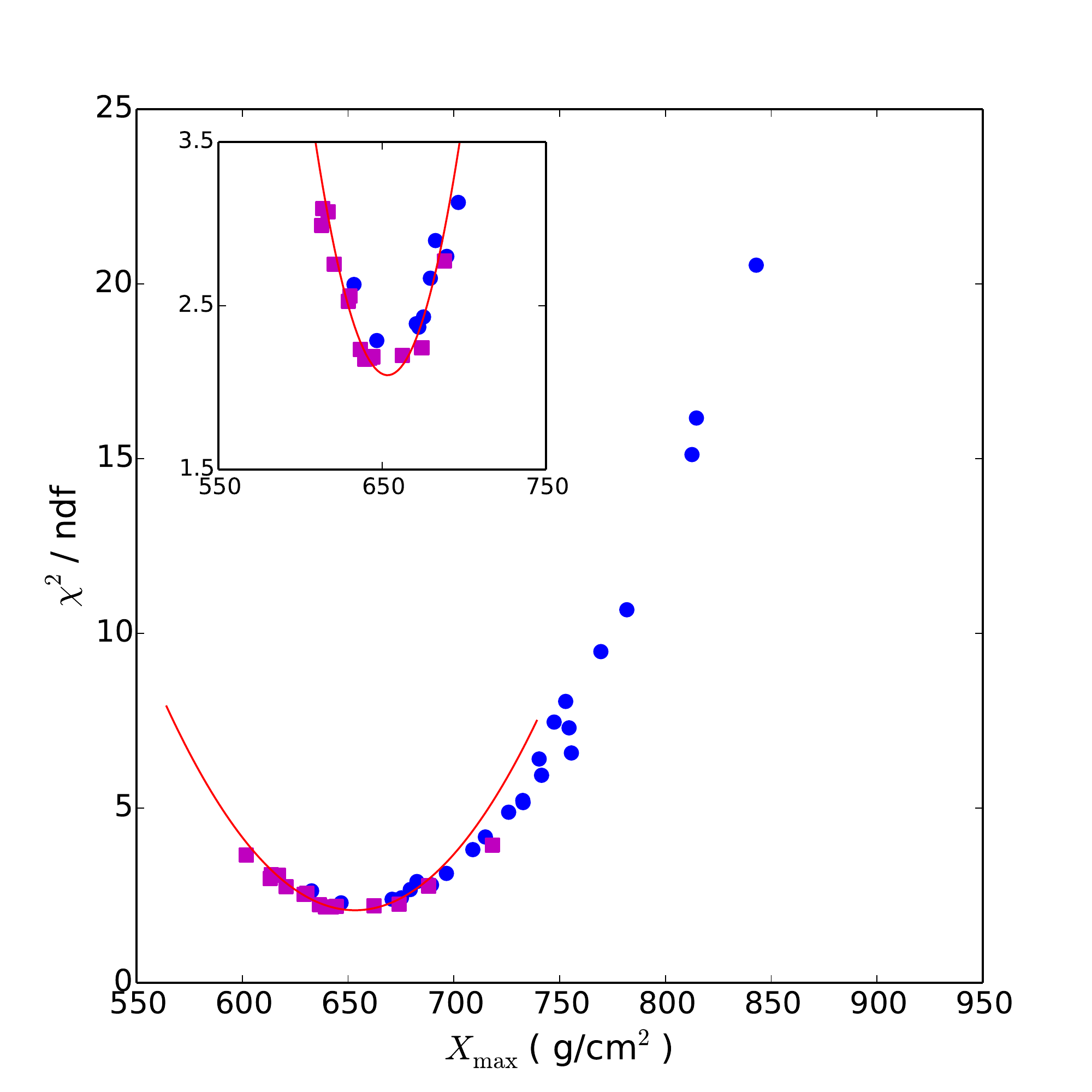}
\includegraphics[width=0.8\linewidth]{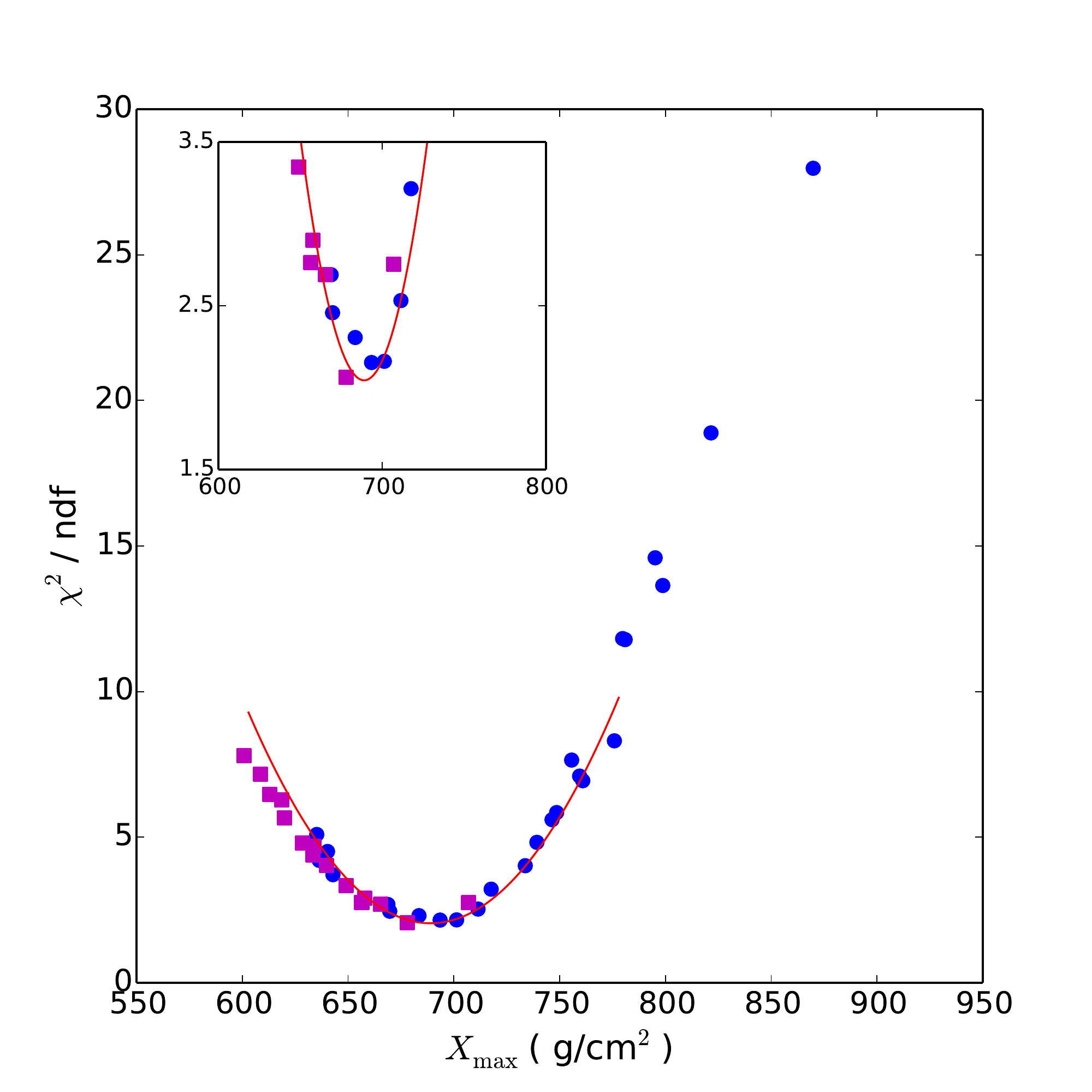}
\caption{\label{fig:xmaxcurves} Reduced $\chi^2$ as a function of $X_\textrm{max}$ for the same three showers as Fig.~\ref{fig:events}. The proton simulations (blue circles) and iron simulations (magenta squares) lie on the same curve, at least within the scatter. A parabola (red line) is fitted to the data points near the minimum to reconstruct $X_\textrm{max}$. The insets zoom in on this region.}
\end{figure}

The fits displayed in Fig.~\ref{fig:events} correspond to the simulation that yielded the lowest $\chi^2$. The reduced $\chi^2$-values for all forty simulations that are performed per detected shower are plotted in Fig.~\ref{fig:xmaxcurves} as a function of the slanted atmospheric depth of the shower maximum $X_\textrm{max}$. While $X_\textrm{max}$ is not the only 
shower parameter that is different between simulations because of shower-to-shower fluctuations,
it is clearly the parameter that most strongly determines the quality of the fit. However, smaller effects due to other variations in the shower development introduce a `jitter'. It is therefore not expected that the data points in Fig.~\ref{fig:xmaxcurves} lie on a completely smooth curve.

The blue circles represent proton simulations and the magenta squares stand for iron simulations. Proton showers on average penetrate deeper into the atmosphere than iron showers and have larger shower-to-shower fluctuations. Indeed, the proton showers cover a larger range of higher $X_\textrm{max}$-values than the iron showers. Interestingly, in the region of overlap the data points of the different primaries follow the same curve, at least within the uncertainty of the above-mentioned jitter. We therefore conclude that showers with the same $X_\textrm{max}$ produce a very similar radiation pattern regardless of the mass of the cosmic-ray primary.

We fit a parabola to the data points within a 200~g/cm$^2$ range centered around the best-fitting simulation and regard its extremum as the reconstructed value for $X_\textrm{max}$. The uncertainty on this reconstructed value is determined with a Monte Carlo study (see next Section) and is different for each shower. It is tempting to derive the uncertainty from the width of the fitted parabola. However, this is only possible if the data points really follow a smooth curve. The jitter on the $\chi^2$-values introduced by shower-to-shower fluctuations affects the shape of the parabola, and therefore renders it impossible to use the width of the parabola to find the uncertainty.
 
\subsection{Uncertainty on Xmax}
\label{sec:uncerX}
For each measured shower the uncertainty on the reconstructed value for $X_\textrm{max}$ is found by applying the following procedure to the set of simulated showers. First, one simulation is singled out and `fake' data is produced by evaluating the radio map at the position of each LOFAR antenna and adding Gaussian noise of the same level as found in the original data. For the position of each LORA particle detector the total deposited energy as simulated with GEANT4 is determined, and again appropriate noise is added to the signal. Then, the remaining thirty-nine simulated showers are fitted to the fake data set using the same fitting procedure as described in Sec.~\ref{sec:technique}. This yields a value $X_\textrm{reco}$ that can be compared to the actual $X_\textrm{real}$ of the simulated shower. Finally, the procedure is repeated for all forty simulated showers (each time taking care that the simulation that is used to produce the fake data set is excluded from the set of simulations that is used for reconstruction).

\begin{figure}
\includegraphics[width=\linewidth]{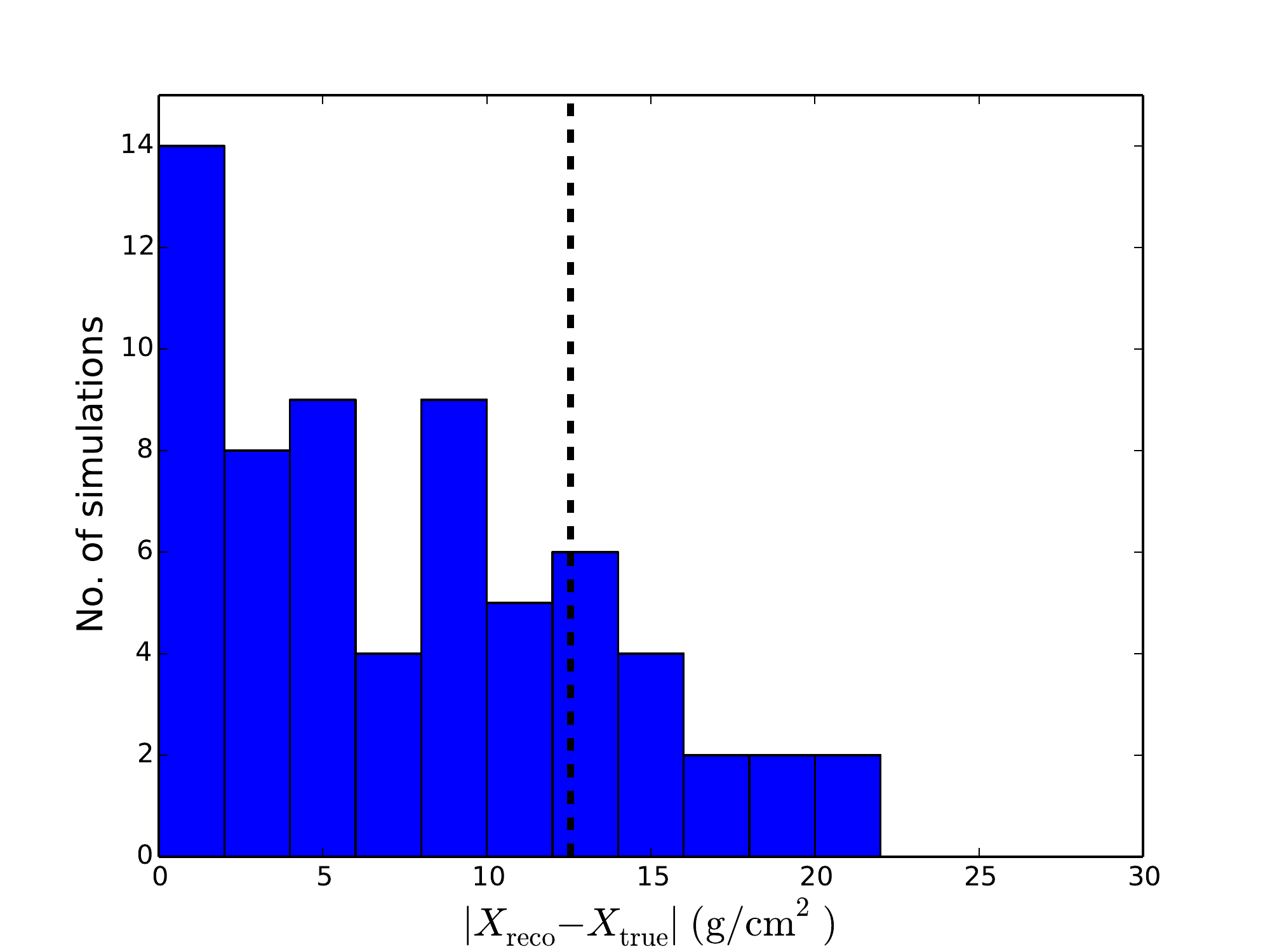}
\caption{\label{fig:uncer1} The uncertainty on $X_\textrm{max}$ for a particular shower is found by reconstructing many simulated showers and evaluating the histogram of the reconstruction error shown here. The black dotted line indicates the value that contains 68\% of the histogram and is taken as the one sigma uncertainty on $X_\textrm{max}$.}
 \end{figure}
 
Fig.~\ref{fig:uncer1} shows the distribution of the $|X_\textrm{reco}-X_\textrm{real}|$ for the forty simulations corresponding to one particular shower. We define the 1$\sigma$ uncertainty as the value of  $|X_\textrm{reco}-X_\textrm{real}|$ that contains 68\% of the histogram. The uncertainty that is found this way is the uncertainty due to the method $\sigma_\textrm{meth}$. There is an additional uncertainty due to the fact that the density profile of the atmosphere at the time of measurement is different from the atmospheric profile used in the CORSIKA simulation. 

To correct for the atmosphere we extract the local atmospheric density profile at the time of measurement from the Global Data Assimilation System (GDAS) \cite{GDAS}. We follow the approach that is used by the Pierre Auger collaboration as described in Abreu et al. \cite{GDASauger}
This work also contains comparisons of atmospheric depth profiles predicted by GDAS and in situ measurements with weather balloons. The differences are typically smaller than 1 g/cm$^2$, except for altitudes very close to the ground. Since global atmospheric models typically work better in the Northern hemisphere where more weather data is available, and the geography of the northern Netherlands is rather unspectacular, we assume that the uncertainty introduced by the atmospheric model is also not worse than 1 g/cm$^2$ at the LOFAR site.

Because the reconstruction of $X_\textrm{max}$ based on the radio emission profile is essentially a geometrical technique, simulations that are produced with a standard atmosphere yield the correct geometrical altitude $h$. The corresponding atmospheric depth is now found by evaluating:
\begin{equation}
X(h)=\frac{1}{\cos\theta}\int^{\infty}_{h} \rho_{\mathrm{GDAS}}(h^{\prime}) dh^{\prime},
\end{equation}
where $\rho_{\mathrm{GDAS}}$ is the atmospheric density profile as predicted by GDAS, and $\theta$ is the zenith angle of the shower. Correction are typically of the order of $\sim 10$~g/cm$^2$.

A third contribution to the uncertainty on $X_\textrm{max}$ comes from the uncertainty in the direction reconstruction. In this analysis, we have used a plane wave approximation which has an angular resolution of $\sim 1^{\circ}$, which translates into an uncertainty of $\sim2$ g/cm$^2$ depending on zenith angle and shower depth. Using a more realistic reconstruction based on hyperbolic wavefront shapes, the accuracy increases to $\sim 0.1^{\circ}$ \cite{corstanje2014}.



\begin{figure}
\includegraphics[width=\linewidth]{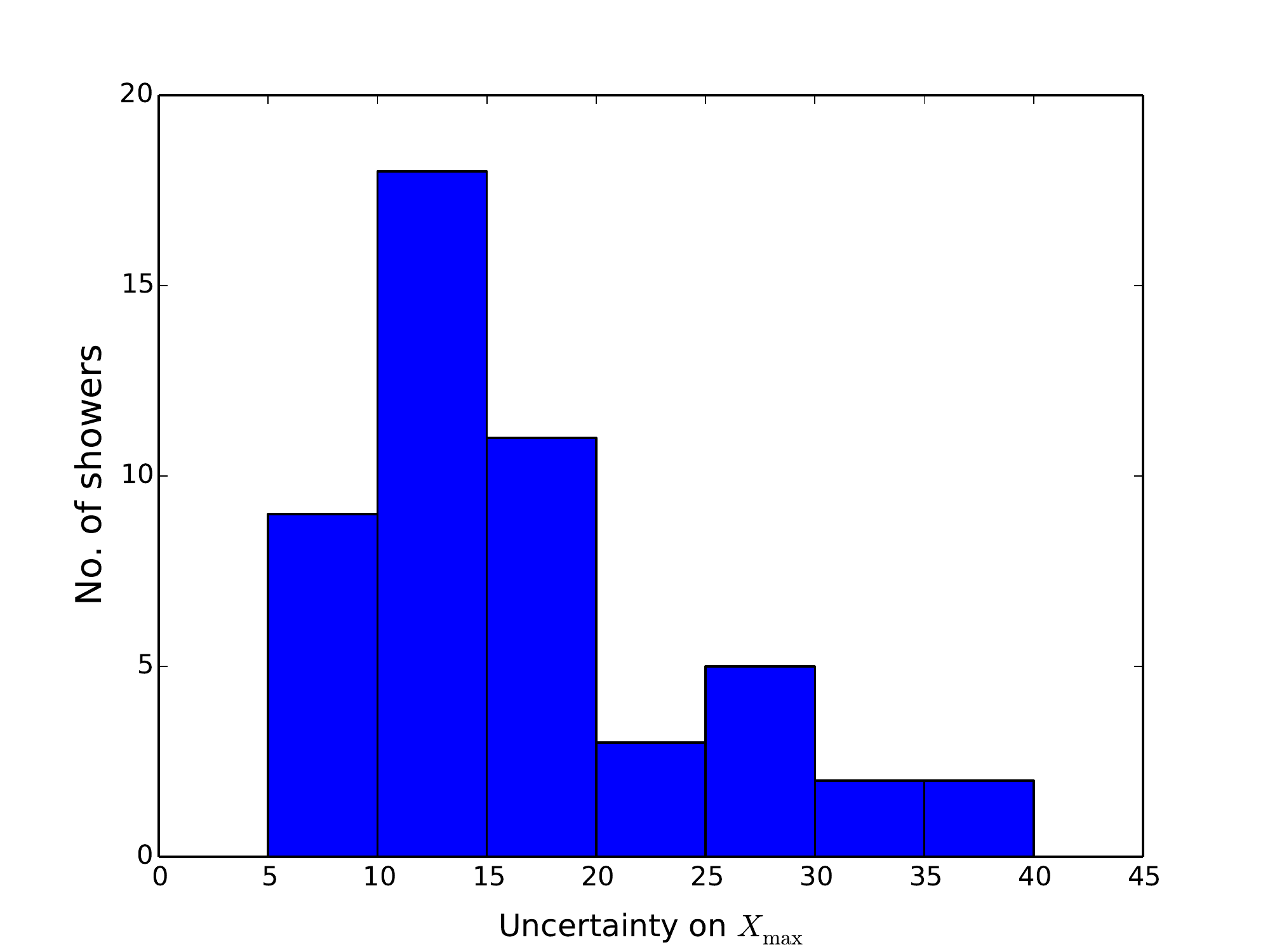}
\caption{\label{fig:uncer2} Histogram of uncertainties on $X_\textrm{max}$ for fifty sets of simulated LOFAR showers. The mean value is 17 g/cm$^{2}$.}
\end{figure}

Simulation sets were generated for fifty showers (each set consisting of 25 proton and 15 iron showers). The uncertainties on $X_\textrm{max}$ for these showers, as has been evaluated with the technique described above, are plotted in the histogram in Fig.~\ref{fig:uncer2}. They range from 7.5 to 37.5 g/cm$^2$, depending on the geometry of the event, with a mean value of 17 g/cm$^{2}$.

\section{Systematic effects}
\label{sec:systematics}
In this Section we study the possible systematic effects that are introduced by the reconstruction method and the choice of models used for simulation.
\subsection{Multivariate fit procedure}
The shower simulations are fitted to the data using Eqn.~\ref{eq:fit}, which has four free parameters: two for the core location, one scaling factor for the radio power $f_{r}^2$ and one scaling factor for the deposited energy in the particle detectors $f_{p}$. A multivariate fit can introduce systematic biases in one or more of the fit parameters. We study this using the same approach as described in Sec.~\ref{sec:uncerX}. Each simulated shower is used to construct a fake data set which is reconstructed using the thirty-nine remaining simulations made for that particular shower. For each reconstruction, the fit parameters are compared to the actual values of the simulated event. This is done for a total of fifty showers.

\begin{figure*}
\includegraphics[width=0.45\linewidth]{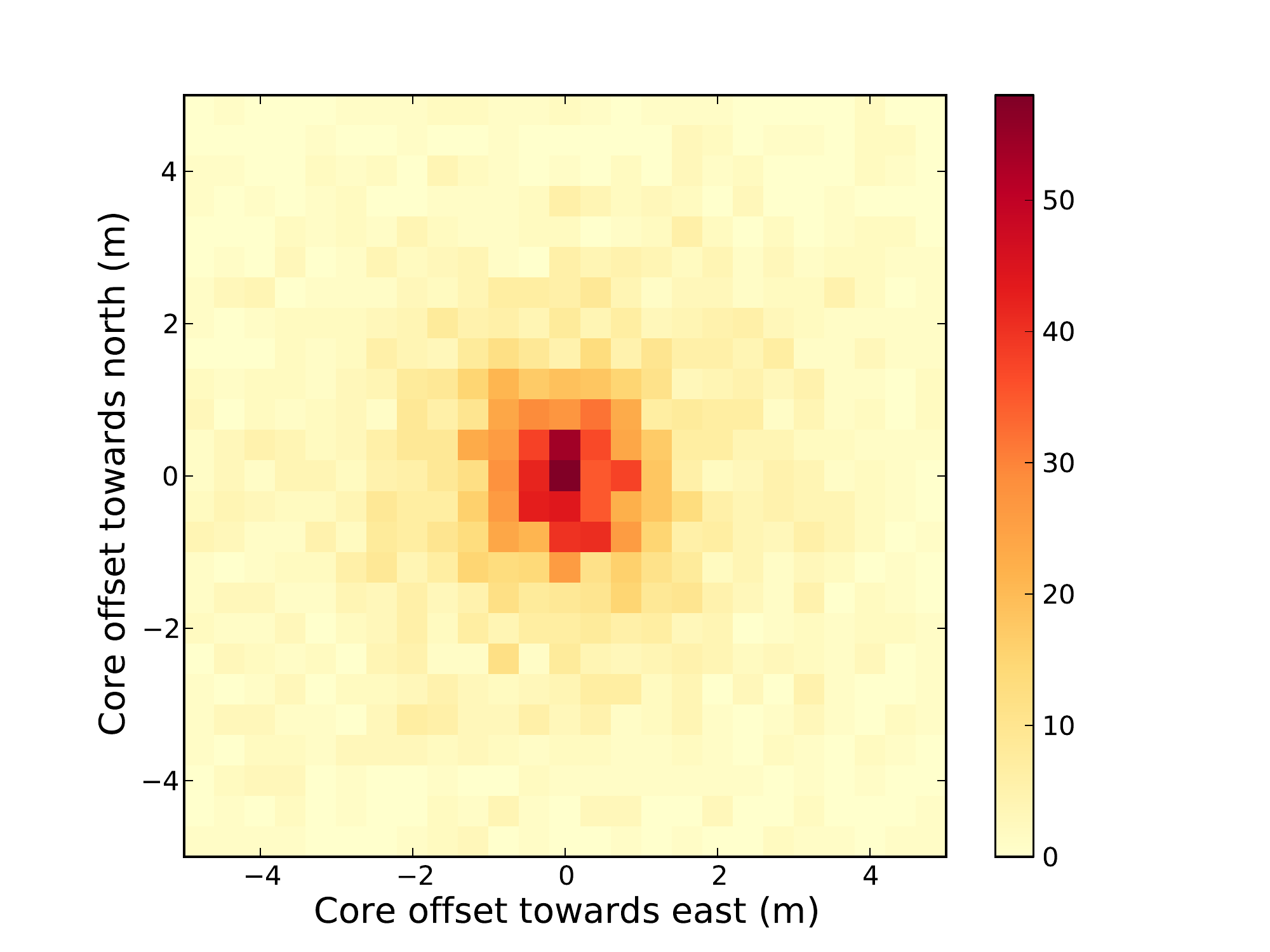}
\includegraphics[width=0.45\linewidth]{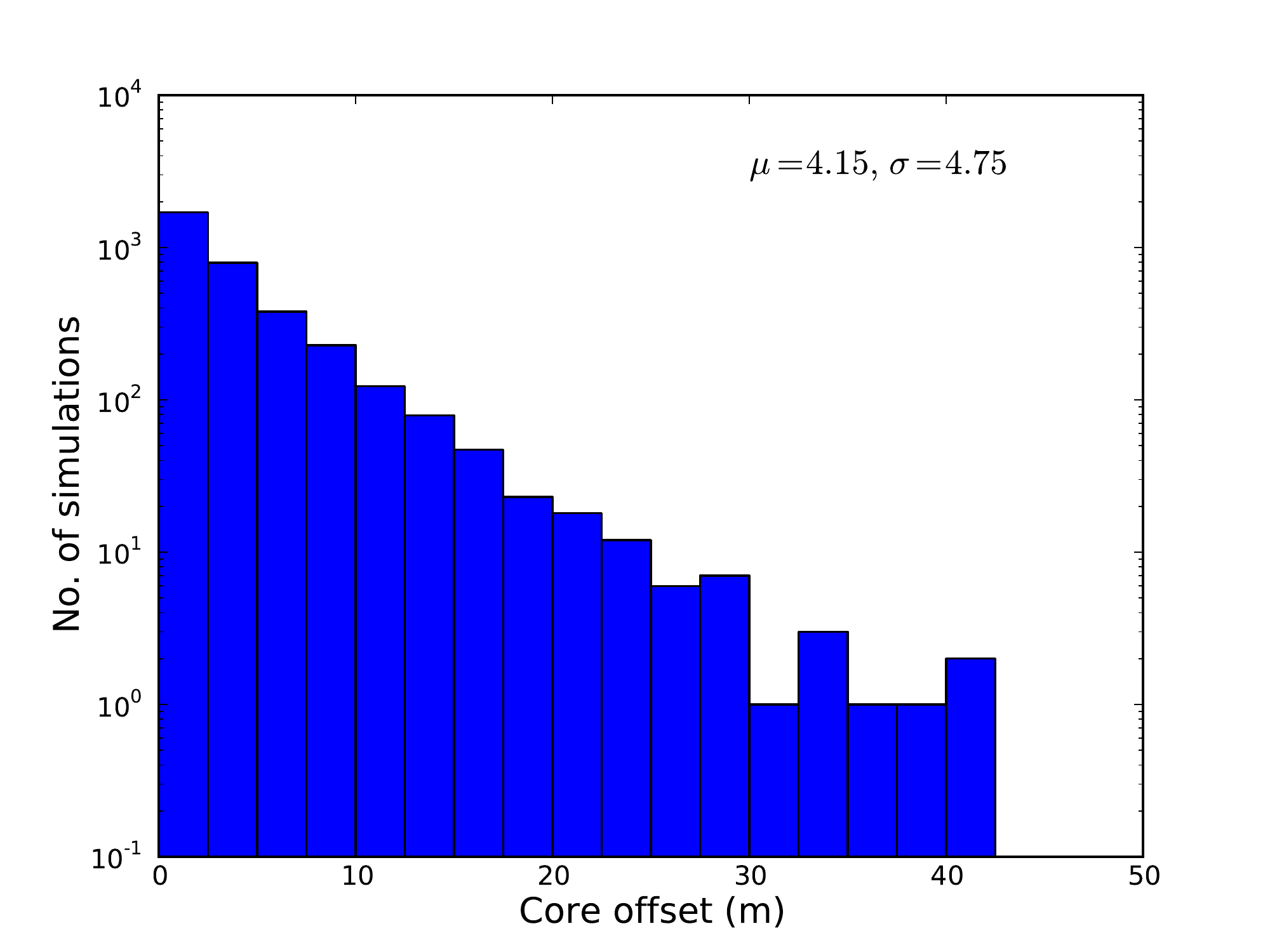}

\includegraphics[width=0.45\linewidth]{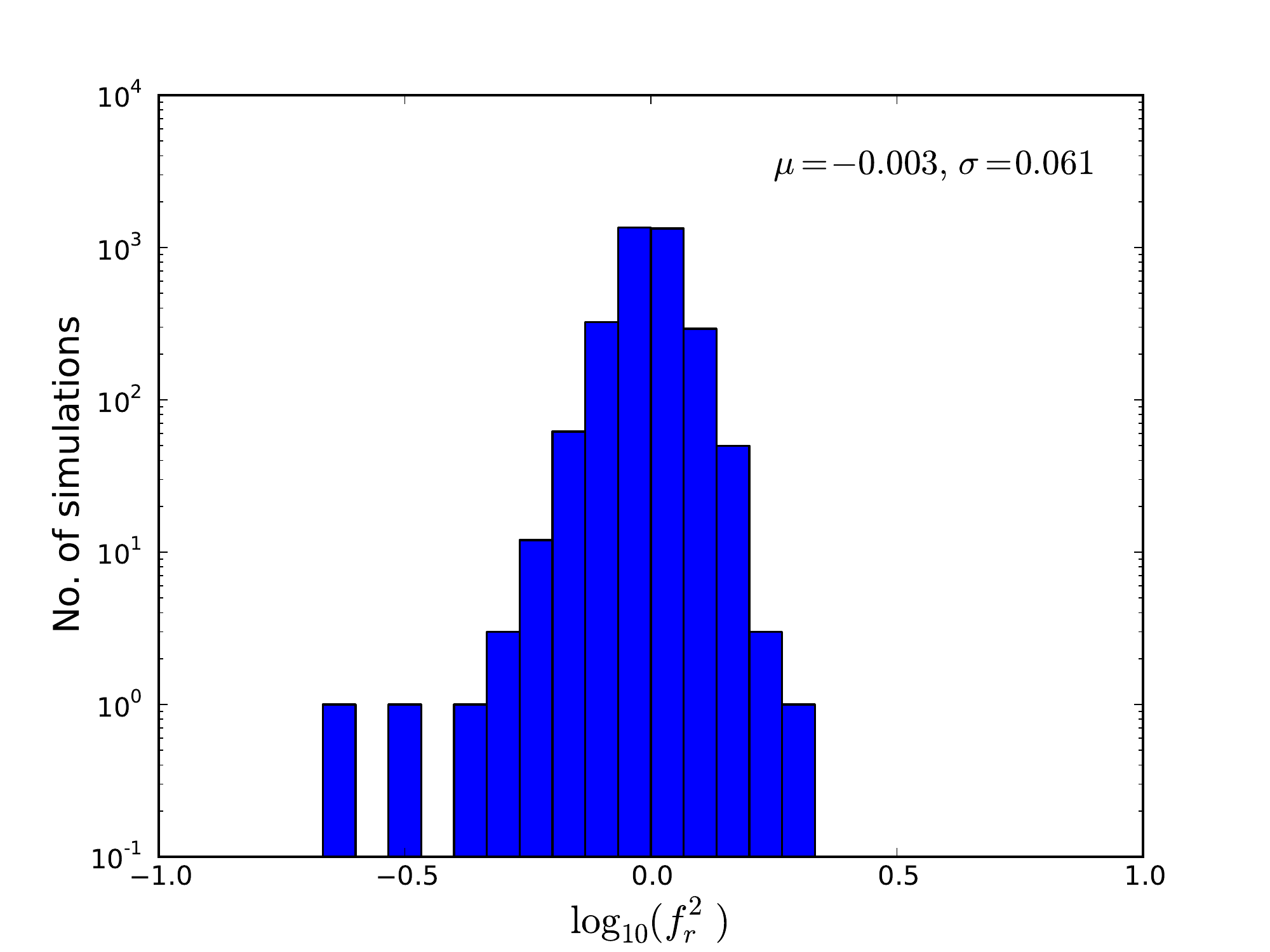}
\includegraphics[width=0.45\linewidth]{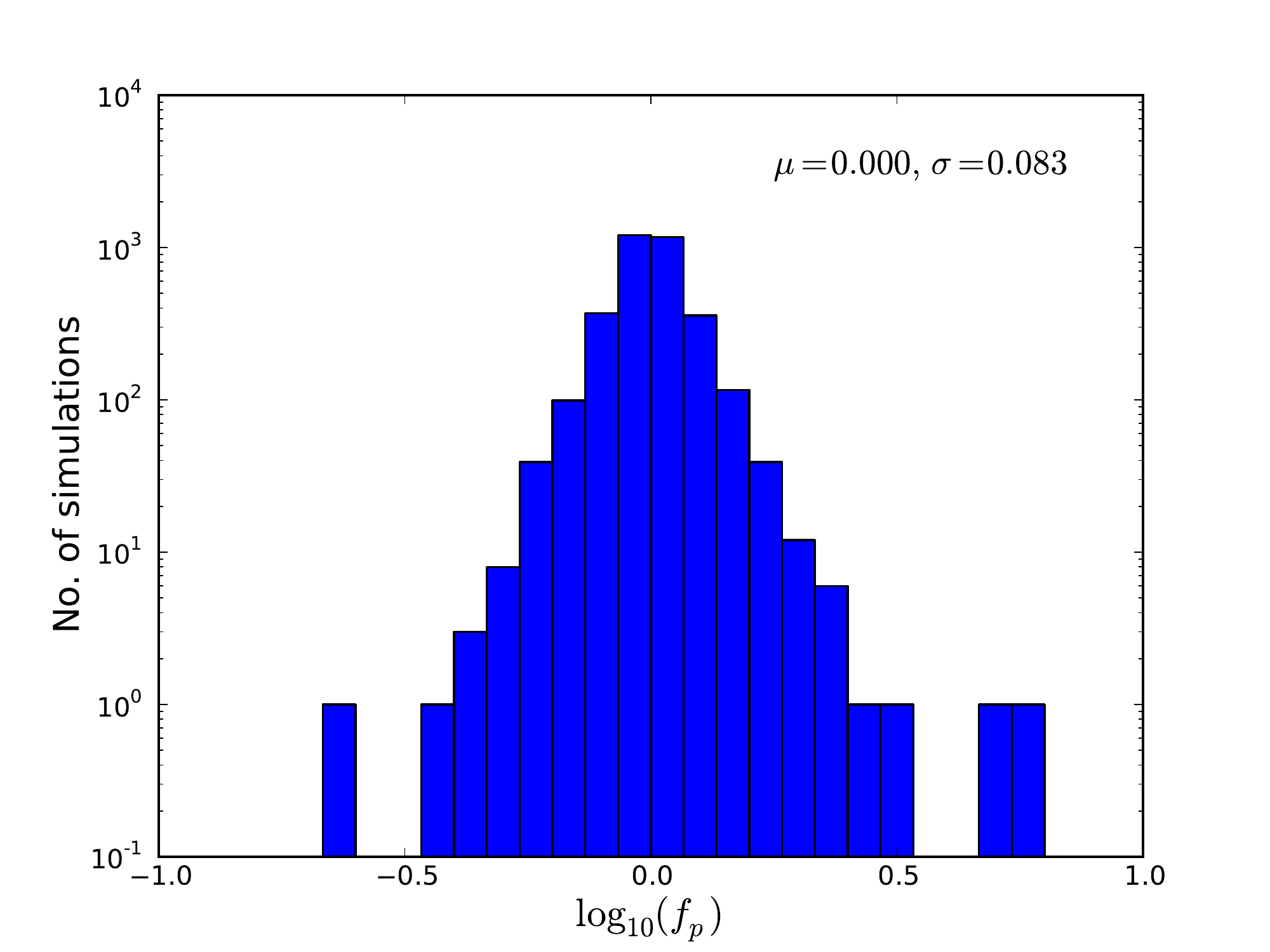}

\caption{\label{fig:fitparamhists} Histograms of fitting parameters for fifty showers with forty simulations each. Top row: distribution of core offset in one and two dimensions. Bottom row: distribution of the logarithm of the scaling parameters $f_r^2$ and $f_p$.}
 \end{figure*}

The results are shown in Fig.~\ref{fig:fitparamhists}. The top two panels show the offset of the reconstructed core position with respect to the real core position. The left panel is a two-dimensional histogram of the core offset in which it can be seen that the core offset has no preferred direction. Hence, there is no systematic effect on the core position due to the fit procedure. The absolute value of the core offset is histogrammed in the right hand panel. The core position is reconstructed with an accuracy of within 5~m.

The bottom panels of Fig.~\ref{fig:fitparamhists} display the distribution of the logarithm of the scaling parameters $f_{p}$ and $f_{r}^2$. Since all forty simulations of a specific shower have the same primary energy, both factors are unity when the reconstruction is perfect. Indeed, the histograms of both scaling factors are symmetric around unity. The maximum energy resolution that can be achieved with this method is given by the width of the distributions and is 15-20\%. The resolution of the radio energy scaling is slightly better than the particle energy scaling. 
  
\subsection{Choice of hadronic interaction model}
\label{hadronic}
The longitudinal development of air showers is sensitive to high-energy hadronic interactions. Hadronic cross-sections, elasticities, and multiplicities cannot be calculated from first principles.
Instead, phenomenological hadronic interaction models are used which are fine-tuned to available accelerator data, but extrapolated to regimes in energy and phase-space far beyond the reach of any Earth-based accelerator \cite{Ulrich2011}

Differences in high-energy cross sections between models result in systematically different values for $X_\textrm{max}$ given a certain primary mass and energy. For example, the difference in the mean atmospheric depth of the shower maximum for proton primaries as predicted by QGSJETII-04 and EPOS-LHC is of the order of 20 g/cm$^2$ at 10$^{18}$~eV \cite{Pierog2013}. Since the measurement of $X_\textrm{max}$ using the radio profile is a geometrical measurement (like fluorescence measurements), it can be argued that there is no systematic effect on the reconstructed depth due to the choice of hadronic model, and that this choice only becomes essential when interpreting the data, i.e.\ when deriving primary mass composition from $X_\textrm{max}$ measurements.

However, it is possible that the reconstructed value of $X_\textrm{max}$ systematically shifts when the shapes of the longitudinal development of actual showers is different from those of the simulated showers. To evaluate this effect, we have generated shower simulations based on EPOS-LHC and SIBYLL 2.1 for ten showers, and reconstructed their $X_\textrm{max}$ values using QGSJETII-04 simulations.

\begin{figure}
\includegraphics[width=0.9\linewidth]{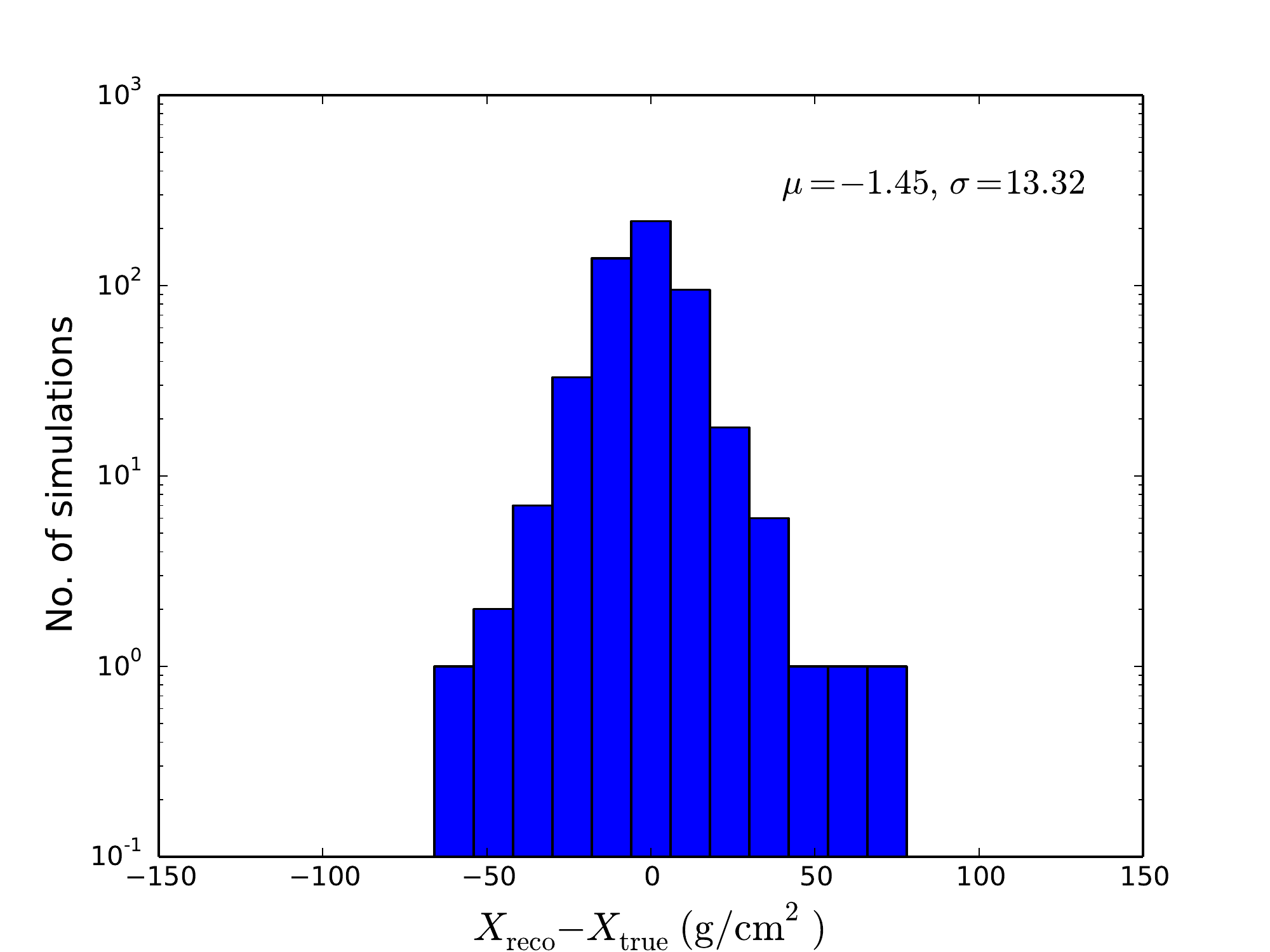}
\includegraphics[width=0.9\linewidth]{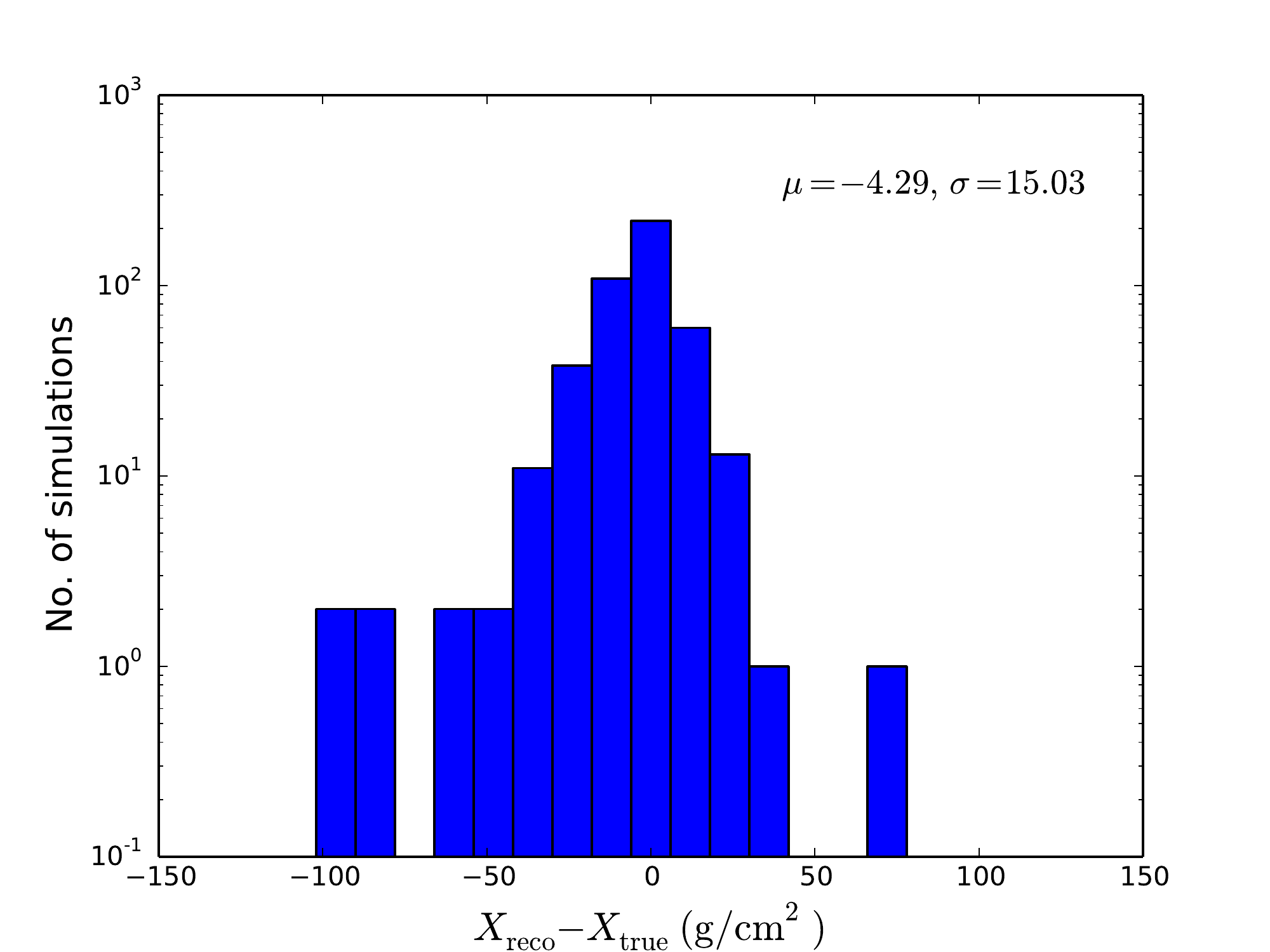}
\includegraphics[width=0.9\linewidth]{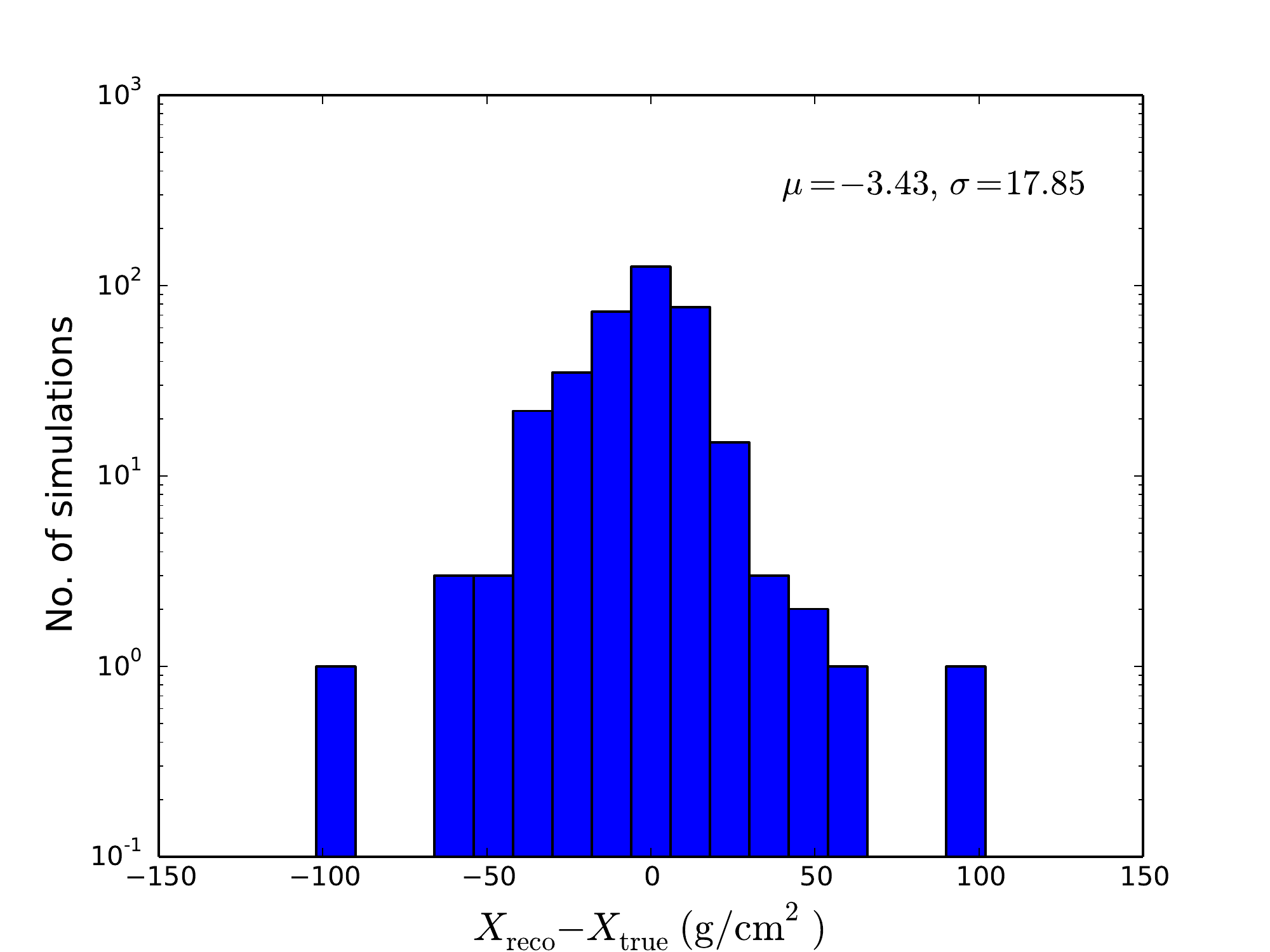}
\caption{\label{fig:hadrsyst} Distribution of the error on the reconstructed value of $X_\textrm{max}$ for simulations based on different hadronic interaction models (top: QGSJETII.04, middle: EPOS-LHC, bottom: SIBYLL 2.1). In each case the reconstruction was done with a sample of QGSJETII.04 showers.}
 \end{figure}

The results are shown in Fig.~\ref{fig:hadrsyst}. The largest systematic offset of 4.3 g/cm$^2$ is found for showers simulated with EPOS-LHC. Note, however, that current experimental constraints on hadronic interactions may very well allow parameter values that produce larger differences than those observed between these three particular models. 

\subsection{Choice of radio simulation code}
\label{sec:radiocode}
There are currently four advanced codes that calculate the radio emission from air showers. CoREAS and ZHAireS \cite{zhaires} are both microscopic codes, i.e.\ they sum the contribution of individual electrons and positrons to obtain the total radio pulse. CoREAS is based on CORSIKA and the end-point formalism, while ZHAireS is based on Aires and the ZHS algorithm \cite{ZHS}. Still, both codes produce very similar results \cite{ModelConvergence}.

Selfas2 \cite{selfas} and EVA \cite{eva} follow a more macroscopic approach, in which shower simulation codes are used to obtain histograms or parameterizations of the total charge distribution of the shower. The radiation is then calculated in a second step. A disadvantage of this approach is that the use of histograms or parameterizations may lead to the loss of information about the charge distribution. An advantage, however, is that such codes can provide a better qualitative understanding of the radiation mechanism. They can, for example, be used to calculate specific components of the total radiation (geomagnetic or charge excess), whereas the microscopic approach is oblivious to such distinctions as it calculates the radiation from first principles.    

A detailed comparison of these four codes with LOFAR data is currently being made and will be the subject of a future publication. For now, we emphasize that for the example events shown here, the CoREAS simulations reproduce all features observed in the data and are able to fit the data with excellent reduced $\chi^2$-values, and that CoREAS and ZHAireS produce very comparable results.

Finally, it should be understood that the correct simulation of the radio signal is mainly a numerical challenge, since the laws of electrodynamics are well-known. This is in sharp contrast with the uncertainties introduced by the high-energy hadronic interactions, which can currently not be derived from first principles.  

\section{Conclusion}
\label{sec:conclusion}
We have presented a new method to reconstruct the atmospheric depth of the shower maximum $X_\textrm{max}$ with radio measurements. It is based on the complete two-dimensional distribution of the emitted radio power which strongly depends on the longitudinal development of the shower. Application of the technique to LOFAR data yields very accurate reconstructions of typically 17 g/cm$^2$. This makes LOFAR an excellent observatory to study the cosmic-ray composition in the energy regime of 10$^{17}$-10$^{18}$~eV, which may harbour the transition from a Galactic to extragalactic origin.  

The radiation profiles that are produced with the CoREAS radio simulation code fit the data extremely well. All features in the complicated, asymmetric profiles are reproduced and we find low reduced $\chi^2$-values for showers that were observed with hundreds of antennas simultaneously. This inspires confidence that the radiation mechanism is now well-understood and can be accurately simulated. The performance of other radio simulation codes is currently being studied and will be published separately.

We have followed a hybrid approach that combines the total radio power and particle measurements in a single fit. Because of the large number of antennas, the radio data give the dominant contribution to the fit of the shower core position. However, by including the particle detector data in the fit we have made sure that the shower reconstruction is fully consistent with all available data. 

It is possible to extend the technique to incorporate information of the radio pulse that is currently not used. The polarization and spectrum of the pulse both depend on the antenna position relative to the shower axis, and pulse arrival times can be used to fit the shower front shape leading to a higher angular accuracy.    

Producing large sets of radio simulations for each detected shower requires a large amount of computational resources. The process can be streamlined by making use of a two-dimensional parameterization of the radiation profile \cite{NellesLDF, AM2014}. 
 
Radio detection of air showers provides a new way of accurately measuring $X_\textrm{max}$. In contrast to fluorescence detection it has a duty cycle of nearly 100\% and may therefore be an interesting method for cosmic-ray composition studies at the highest energies. The main challenge lies in the size of the radio footprint, which is smaller than the particle footprint and requires a relatively dense antenna array. The technique itself, however, has matured and now produces accurate and robust results.   
\\

\begin{acknowledgments}
We acknowledge financial support from  the Netherlands Organization for Scientific Research (NWO), VENI grant 639-041-130, the Netherlands Research School for Astronomy (NOVA), the Samenwerkingsverband Noord-Nederland (SNN) and the Foundation for Fundamental Research on Matter (FOM). We acknowledge funding from an Advanced Grant of the European Research Council under the European Unions Seventh Framework Program
(FP/2007-2013) / ERC Grant Agreement n.~227610. Part of this work was supported by grant number VH-NG-413 of the Helmholtz Association.
LOFAR, the Low Frequency Array designed and constructed by ASTRON, has facilities in several countries, that are owned by various parties (each with their own funding sources), and that are collectively operated by the International LOFAR Telescope (ILT) foundation under a joint scientific policy.
\end{acknowledgments}


\end{document}